\documentclass[aps,showpacs,superscriptaddress,preprintnumbers,nofootinbib]{revtex4}
\usepackage{slashed,color,amsmath,amssymb,mathrsfs}
\usepackage{graphicx}
\usepackage{hyperref}
\usepackage{longtable}
\usepackage{array}

\allowdisplaybreaks
 \def\zh{\tiny}

 \def\la{\langle}
 \def\ra{\rangle}

 \def\chip{\chi_+}
 
 \def\chim{\chi_-}
 
 \def\fp{f_+}
 \def\fm{f_-}
 \def\tb{\bar{t}}
 \def\td{{t^{\dag}}}
 \def\tp{t_+}
 \def\tm{t_-}

 \def\FL{F_L}
 \def\FR{F_R}

 \def\chid{{\chi^\dag}}

 \def\ud{{u^\dag}}
 \def\xu{u}
 \def\thetah{\hat{\theta}}
 \def\detx{\det\chi}
 \def\detxd{\det\chid}

 \def\Gam{\Gamma}
 \def\Gam{\Gamma}
 \def\Ud{U^\dag}
\begin{document}
\preprint{}
\title{Full pseudoscalar mesonic chiral Lagrangian at $p^6$ order under the unitary group}
\author{Shao-Zhou Jiang}
\email[]{jsz@gxu.edu.cn}
\affiliation{Department of Physics and GXU-NAOC Center for Astrophysics and Space Sciences, Guangxi University,
Nanning, Guangxi 530004, P.R.China}

\author{Feng-Jun Ge}
\email[]{ge\_fengjun@iapcm.ac.cn}
\affiliation{Institute of Applied Physics and Computational Mathematics, Beijing, 100094, P.R.China}

\author{Qing Wang}
\email[Corresponding author.]{wangq@mail.tsinghua.edu.cn}
\affiliation{Department of Physics, Tsinghua University, Beijing 100084, P.R.China}
\affiliation{Center for High Energy Physics, Tsinghua University, Beijing 100084, P.R.China}
\affiliation{Collaborative Innovation Center of Quantum Matter, Beijing 100084, P.R.China}

 \begin{abstract}

 We construct the full $p^6$ order chiral Lagrangians for the unitary group and special unitary groups,
 including $n_f$-, three- and two-flavor cases, all bilinear currents
 (scalar, pseudoscalar, vector, axial-vector and tensor currents) and $\theta$ parameter.
 The number of independent operators are 1391, 1326 and 969 for each of the flavor unitary groups.
 From these results, we find one extra linear relation among the traditional $p^4$ order low-energy constants under the U(3) group,
 and some more linear relations with tensor sources for the $p^6$ order low-energy constants
 in the special unitary groups.
 We develop a scheme to obtain the relations for the dependent operators in terms of independent operators.

 \end{abstract}
 \maketitle
\section{Introduction}
 In low-energy QCD, chiral perturbation theory(ChPT) is a powerful tool to treat hadron physics.
 With the help of ChPT, we can describe the low-energy pseudoscalar mesons ($\pi,K,\eta,\eta'$) up to a certain degree of precision.
 In the last three decades, ChPT has matured and can specify next-to-next-to-leading order (NNLO) processes.
 The first step in the ChPT is to obtain the chiral Lagrangian, where most of the difficulties and discussions arise.
 Conventionally, one expands the chiral Lagrangian in terms of powers of  momentum ($p$).
 For the special unitary (SU) group, ChPT for the pseudoscalar meson has been improved from the leading ($p^2$) order \cite{weinberg},
 to the next-to-leading ($p^4$) order (NLO) \cite{GS1,GS2},
 and the NNLO ($p^6$ order) \cite{p61,p62,p6p,p6a1,p6a2,tensor1,ourt}.
 At present, almost all NNLO chiral Lagrangian have been obtained, including two- and three-flavor quarks,
 the normal and anomalous parts,
 and all bilinear light-quark currents (scalar, pseudoscalar, vector, axial-vector and tensor currents),
 except for the $\theta$ parameter related terms.
 For the unitary (U) groups, the NLO results also were obtained\cite{U3}.
 Furthermore, one can expand the chiral Lagrangian in terms of $p$ and $1/N_c$ simultaneously.
 The results to order $O(\delta)$ have been obtained \cite{Nc1,Nc1p,Nc2}.

 At present, under the SU group, the NNLO chiral Lagrangian seems sufficient.
 The chiral Lagrangian at any higher order is much more complicated and without much physical interest.
 Even though it can be obtained, the work would be very tedious and long, because of the complexity,
 and the advantages of such a chiral Lagrangian would vanish. Furthermore, comparisons between theory
 and experiments have not reached the precision where higher-order computations are needed.
 Even in the NNLO, some mistakes may appear through the tedium of the calculations,
 raising doubts about the credibility of the evaluations. When the NNLO chiral Lagrangian was first obtained \cite{p61,tensor1},
 some linear relations among low-energy constants (LECs) had been missed \cite{p62,p6p,p6a1,p6a2,ourt}.
 This poses the question: Do other hidden relations exit?
 We need a definite answer. If the existing LECs are not independent, they are not unique,
 and ChPT cannot work well. In the NNLO,
 when ones wants all LECs \cite{our5,oura2,ourt}
 or discusses some processes \cite{pp1,pp2,tpf,pk,sf,etp},
 one may need these linear relations among the independent and dependent operators.
 Some relations appearing in the literature are quite long (see \cite{p6p}, Appendix B in \cite{p62} and so on),
 How are these complex calculations to be done, and computations made more reliable?

 Besides more higher-order terms, or remaining with the NNLO under the SU groups,
 one may extend the calculation to the NNLO under the U group.
 For  the conventional SU chiral Lagrangian at the NNLO,
 there are singlet vector sources, singlet axial-vector sources
 or $\theta$ parameter, but sometimes we need to consider them.
 Even though in NLO, these circumstances may be consider, but not very accurate comparing to NNLO.
 The extra $U(1)$ symmetry is related to the $\eta'$ particle. The properties of $\eta'$ \cite{etp1},
 especially $\eta-\eta'$ mixing \cite{etet1,etet2,etet3,etet4,etet5}, need this $U(1)$ symmetry.
 Parameter $\theta$ comes from strong CP violation \cite{theta1,theta2},
 whereas certain phenomenons are related to singlet vector or axial-vector sources, for instance,
 electroweak interactions \cite{ele1,ele2,ele3}. In these situation,
 $\theta$, singlet vector and axial-vector sources cannot be ignored.
 In the $1/N_c$ expansion, $\eta'$ could also be included \cite{Nc1,Nc2,Nc3}.
 Sometimes, tensor sources are needed; for example, the magnetic susceptibility of the quark condensate\cite{e3,e1,e2}.
 The purpose of our study is to obtain the full U group chiral Lagrangian  up to NNLO,
 including all bilinear quark currents and $\theta$ parameters.
 With this chiral Lagrangian and techniques developed,
 all the above mentioned problems can be solved more accurately and simultaneously.

 This paper is organized as follows: In Sec. \ref{symd}, we review the two kinds of basis to construct the chiral Lagrangian
 and provide definitions of symbols. In Sec. \ref{lico}, we collect all possible linear relations for the chiral Lagrangian.
 Sec. \ref{consr} is an introduction to our method of generating the chiral Lagrangian.
 In Sec. \ref{res}, we list our results for the U and SU groups up to order $p^6$.
 Sec. \ref{summ} concludes with a summary.

\section{Chiral Transformations and Chiral Basis}\label{symd}
 In QCD, the original Lagrangian is denoted $\mathscr{L}^0_{\mathrm{QCD}}$.
 We usually introduce some external sources and the topological charge operator to obtain the corresponding Green's functions.
 The complete bilinear coupling external sources contain scalar ($s$), pseudoscalar ($p$),
 vector ($v^\mu$), axial-vector ($a^\mu$),
 and tensor ($\bar{t}^{\mu\nu}$) currents. The topological charge operator is $G_{\mu\nu}\tilde{G}^{\mu\nu}$,
 where $G^{\mu\nu}$ is the gluon field strength matrix, and $\tilde{G}^{\mu\nu}=\epsilon^{\mu\nu\lambda\rho}G_{\lambda\rho}$.
 The QCD Lagrangian is expanded as
 \begin{eqnarray}
 \mathscr{L}=\mathscr{L}^0_{\mathrm{QCD}}
 +\bar{q}(\slashed{v}(x)+\slashed{a}(x)\gamma_5-s(x)+ip(x)\gamma_5+\sigma_{\mu\nu}\bar{t}^{\mu\nu}(x))q
 -\frac{1}{16\pi^2}\theta(x)\mathrm{tr}_c(G_{\mu\nu}\tilde{G}^{\mu\nu}),\label{lag1}
 \end{eqnarray}
 where $q$  denotes the light quark fields.
 For low-energy QCD, we mainly focus on the nonet of the pseudoscalar mesons $(\pi,K,\eta,\eta')$.
 As in \cite{GS1,GS2}, we collect the nonet in the $U(x)$ matrix as
 \begin{eqnarray}
 U(x)&=&e^{i\phi(x)/F_0}\\
 \phi(x)&&\hspace*{-0.5cm}\stackrel{n_f=3}{====}\sum_{a=0}^8\lambda_a\phi_a(x)=\left(
 \begin{array}{ccc}
 \pi^0+\frac{1}{\sqrt{3}}\eta+\sqrt{\frac{2}{3}}\eta' & \sqrt{2}\pi^+ & \sqrt{2}K^+\\
 \sqrt{2}\pi^- & -\pi^0+\frac{1}{\sqrt{3}}\eta+\sqrt{\frac{2}{3}}\eta' & \sqrt{2}K^0\\
 \sqrt{2}K^- & \sqrt{2}\bar{K}^0 & -\frac{2}{\sqrt{3}}\eta+\sqrt{\frac{2}{3}}\eta'
 \end{array}\right).
 \end{eqnarray}
 Where $F_0$ is the pseudoscalar meson decay constant in the chiral limit,
 and $\lambda_i(i=1\ldots 8)$ are the Gell-Mann matrices, and $\lambda_0=\sqrt{2/n_f}\mathrm{I}_{n_f\times n_f}$.
 We usually choose $n_f=2$ or $3$.
 If the light quark are assumed massless, $\mathcal{L}$ is invariant under the $U_L(n_f)\times U_R(n_f)$ group.
 For an effective field theory,
 we often focus on the meson fields $U$.
 Under the $U_L(n_f)\times U_R(n_f)$ transformations, $U$ transforms as
 \begin{eqnarray}
 U\to g_R U g_L^\dag,\hspace{0.5cm}g_L\in U_L,g_R\in U_R.\label{utrans}
 \end{eqnarray}
 Simultaneously, the external sources also need transforming. For convenience,
 we usually collet the sources as
 \begin{eqnarray}
 &&r^\mu=v^\mu+a^\mu,\notag\\
 &&l^\mu=v^\mu-a^\mu,\notag\\
 &&\chi=2B_0(s+ip),\notag\\
 &&\hat{\theta}=i\theta,\notag\\
 &&X=\la\ln U\ra+\thetah,\notag\\
 &&t^{\mu\nu}=\frac{1}{2}\bar{t}^{\mu\nu}-\frac{i}{4}\epsilon^{\mu\nu\lambda\rho}\bar{t}_{\lambda\rho}.\label{bas}
 \end{eqnarray}
 The notation $\la\cdots\ra$ is the trace over flavor indices.
 $B_0$ is a constant related to the quark condensate.
 Under $U_L(n_f)\times U_R(n_f)$ rotations (chiral rotations), these sources transform as
 \begin{eqnarray}
 &&l^{\mu}\to g_L l_\mu g^\dag_L+ig_L\partial^{\mu}g^\dag_L\notag\\
 &&r^{\mu}\to g_R r_\mu g^\dag_R+ig_R\partial^{\mu}g^\dag_R\notag\\
 &&\chi\to g_R \chi g^\dag_L\notag\\
 &&t^{\mu\nu}\to g_R t^{\mu\nu} g^\dag_L\notag\\
 &&X\to X\label{jbt}
 \end{eqnarray}
 For $\thetah$, one usually introduce its covariant derivative
 \begin{eqnarray}
 \nabla^\mu\thetah=\partial^\mu\thetah+2i\la a^\mu\ra,
 \end{eqnarray}
 which is invariant under the chiral rotations.
 It is more convenient to introduce the field strengths $F_L^{\mu\nu}$ and $F_R^{\mu\nu}$ associated with $l^\mu,r^\mu$,
 \begin{eqnarray}
 &&F_R^{\mu\nu}=\partial^{\mu}r^{\nu}-\partial^{\nu}r^{\mu}-i[r^{\mu},r^{\nu}],\notag\\
 &&F_L^{\mu\nu}=\partial^{\mu}l^{\nu}-\partial^{\nu}l^{\mu}-i[l^{\mu},l^{\nu}],\label{flr}
 \end{eqnarray}
 Under $U_L(n_f)\times U_R(n_f)$ rotations, they transform as
 \begin{eqnarray}
 F_L^{\mu\nu}\to g_L F_L^{\mu\nu}g_L^\dag,\hspace{0.3cm}F_R^{\mu\nu}\to g_R F_R^{\mu\nu}g_R^\dag
 \end{eqnarray}

 We can now construct the chiral Lagrangian, with
 $U,U^\dag,F_L^{\mu\nu},F_R^{\mu\nu},\chi,\chi^\dag,t^{\mu\nu},t^{\dag\mu\nu},\nabla^\mu\thetah$ and $X$,
 as well as their covariant derivatives; we call this set the L(eft)R(ight)-basis.
 These covariant derivations combine with $l^\mu$ and $r^{\mu}$,
 retaining the same transformations as the fields on which they act.
 Different transformations have different covariant derivations.
 They can be obtained from (\ref{jbt}). Following \cite{p61},
 we express them as
 \begin{eqnarray}
 A\to g_R A g^\dag_L &:& D^\mu A=\partial^\mu A-ir^\mu A+iAl^\mu,\notag\\
 B\to g_L B g^\dag_R &:& D^\mu B=\partial^\mu B-il^\mu B+iBr^\mu,\notag\\
 C\to g_R C g^\dag_R &:& D^\mu C=\partial^\mu C-ir^\mu C+iCr^\mu,\notag\\
 D\to g_L D g^\dag_L &:& D^\mu D=\partial^\mu D-il^\mu D+iDl^\mu,\notag\\
 E\to E&:& D^\mu E=\partial^\mu E\label{cod}
 \end{eqnarray}

 In principle, the conditions constraining the chiral Lagrangian relate only to its symmetries,
 which include the Lorentz symmetry, local chiral symmetry, parity symmetry and charge conjugation symmetry.
 We list these properties on for the LR-basis in Table\ref{lrbhb}.
 Because the chiral Lagrangian need to be real, we also list the hermiticity of the LR-basis.
 The operators in Table\ref{lrbhb} with covariant derivations have the same properties.
 \begin{table*}[h]
 \caption{\label{lrbhb}Chiral rotations (R), P, C and hermiticity of the LR-basis (O).}
 \begin{tabular}{ccccc}
 \hline\hline $O$ & $R$ & $P$ & $C$ & h.c. \\
 \hline $U$ & $g_R U g_L^\dag$ & $U^\dag$ & $U^T$ & $U^\dag$\\
 $\chi$ & $g_R\chi g_L^\dag$ & $\chid$ & $\chi^T$ & $\chid$\\
 $\FL^{\mu\nu}$ & $g_L\FL^{\mu\nu}g_L^\dag$ & $\FR^{\mu\nu}$ & $-(\FR^{\mu\nu})^T$ & $\FL^{\mu\nu}$\\
 $\FR^{\mu\nu}$ & $g_R\FR^{\mu\nu}g_R^\dag$ & $\FL^{\mu\nu}$ &  $-(\FL^{\mu\nu})^T$ & $\FR^{\mu\nu}$\\
 $t^{\mu\nu}$ & $g_R t^{\mu\nu}g_L^\dag$ & $t^{\dag\mu\nu}$ & $-(t^{\mu\nu})^T$ & $\td^{\mu\nu}$\\
 $X$ & $X$ & $-X$ & $X$ & $-X$\\
 $\nabla^\mu\thetah$ & $\nabla^\mu\thetah$ & $-\nabla^\mu\thetah$ & $\nabla^\mu\thetah$ & $-\nabla^\mu\thetah$\\
 \hline\hline
 \end{tabular}
 \end{table*}

 At present, we can construct the chiral Lagrangian with the LR-basis.
 However, different bases have different chiral rotation properties, particularly, in \eqref{cod},
 the covariant derivatives relations are complex.
 Such differences can yield among the covariant derivatives quit complex properties for some identities,
 for instance the strength tensor relations and the Bianchi identity,
 which will be introduced later in \eqref{Gam0} and \eqref{bi} respectively, in regard to another basis.
 In using the LR-basis some linear relations may get lost \cite{p61}.
 Hence, we usually use another basis, which we refer to as building blocks.
 We will until the very end use these building blocks to construct the chiral Lagrangian,
 but will return to the LR-basis for the contact terms.

 Following \cite{p62,p6a2,tensor1}, we define the Goldstone boson field $u$, $U=u^2$, and introduce a compensator field $h$.
 The field $u$ transforms as
 \begin{eqnarray}
 u\to u=g_R uh^\dag=h ug_L^\dag.
 \end{eqnarray}
 Hence, under chiral rotations, all the building blocks $O$ are invariant or transform as
 \begin{eqnarray}
 O\to hOh^{\dag}.\label{otr}
 \end{eqnarray}
 These building blocks and their relations to the LR-basis are
 \begin{eqnarray}
 u^\mu&=&i\{u^\dag(\partial^\mu-ir^\mu)u-u(\partial^\mu-il^\mu)u^\dag\},\notag\\
 \chi_\pm&=&u^\dag\chi u^\dag\pm u\chi^\dag u,\notag\\
 h^{\mu\nu}&=&\nabla^\mu u^\nu+\nabla^\nu u^\mu,\notag\\
 f_\pm^{\mu\nu}&=&u F_L^{\mu\nu} u^\dag\pm u^\dag F_R^{\mu\nu} u,\notag\\
 t_\pm^{\mu\nu}&=&u^\dag t^{\mu\nu}u^\dag\pm u t^{\mu\nu\dag} u,\notag\\
 \nabla^{\mu}\thetah&=&\partial^{\mu}\thetah+2i\la a^{\mu}\ra,\notag\\
 X&=&\la\ln U\ra+\thetah.\label{df}
 \end{eqnarray}
 We can see that almost all the operators on the left-hand side of \eqref{df} transform as \eqref{otr};
 the exceptions $\nabla^{\mu}\thetah$ and $X$ are invariant.

 Furthermore, we also need to define the covariant derivative for all operators.
 For the operators transforming as \eqref{otr}, the covariant derivatives are
 \begin{eqnarray}
 \nabla^{\mu}O=\partial^\mu O+[\Gamma^\mu,O],\label{cd}
 \end{eqnarray}
 where the chiral connection is
 \begin{eqnarray}
 \Gamma^{\mu}=\frac{1}{2}\{\ud(\partial^\mu-ir^\mu)u+u(\partial^\mu-il^{\mu})\ud\}.
 \end{eqnarray}
 Because $\nabla^{\mu}\thetah$ and $X$ are invariant under the chiral rotations,
 their covariant derivatives are the ordinary partial derivative.
 For convenience, we will use below $\nabla$ instead of $\partial$,
 because the commutator is zero in \eqref{cd}.
 With this covariant derivative, $\fm^{\mu\nu}$ can be simplified as
 \begin{eqnarray}
 f_-^{\mu\nu}=-\nabla^\mu u^{\nu}+\nabla^\nu u^{\mu}.\label{fm}
 \end{eqnarray}
 The field strength tensor $\Gamma_{\mu\nu}$ associated with this covariant derivative is
 \begin{eqnarray}
 &&[\nabla^\mu,\nabla^\nu]O=[\Gamma^{\mu\nu},O],\label{Gam0}\\
 &&\Gamma^{\mu\nu}=\nabla^{\mu}\Gam^{\nu}-\nabla^{\nu}\Gam^{\mu}-[\Gam^{\mu},\Gam^{\nu}]
 =\frac{1}{4}[u^\mu,u^\nu]-\frac{i}{2}f_{+}^{\mu\nu}.\label{Gam1}
 \end{eqnarray}
 The chiral Lagrangian can be constructed by the building blocks in (\ref{df}) and their covariant derivatives
 \footnote{The $\nabla^\mu$ in $\nabla^\mu\hat{\theta}$ is not this covariant derivative.
 However we believe no confusion arises, because they are always combinative.
 Higher-order covariant derivatives in $\nabla^\mu\hat{\theta}$ are equal to the ordinary derivatives or \eqref{cd}.
 Therefore, we persist with this form for the sake of convenience.}.
 As in \cite{p62}, we choose for $\chi_{\pm}$ -a more convenience form for its covariant derivatives;
 \begin{eqnarray}
 \chi_{\pm}^{\mu}=u^\dag D^\mu \chi u^\dag\pm u D^\mu\chid u=\nabla^\mu\chi_{\pm}-\frac{i}{2}\{\chi_{\mp},u^\mu\},\label{chimu}
 \end{eqnarray}
 which has a more simpler relationship in the LR-basis.
 With regard to the LR-basis, we list the properties of the building blocks in Table\ref{blbt}.
 \begin{table*}[h]
 \caption{\label{blbt}P, C and hermiticity of the building blocks.}
 \begin{tabular}{cccc}
 \hline\hline $O$ & $P$ & $C$ & h.c. \\
 \hline $u^{\mu}$ & $-u_{\mu}$  & $(u^{\mu})^T$ & $u^{\mu}$ \\
 $h^{\mu\nu}$ & $-h_{\mu\nu}$  & $(h^{\mu\nu})^T$ & $h^{\mu\nu}$ \\
 $\chi_{\pm}$ & $\pm\chi_{\pm}$  & $(\chi_{\pm})^T$ & $\pm \chi_{\pm}$\\
 $f_{\pm}^{\mu\nu}$ & $\pm f_{\pm\mu\nu}$  & $\mp (f_{\pm}^{\mu\nu})^T$ & $ f_{\pm}^{\mu\nu}$  \\
 $t_{\pm}^{\mu\nu}$ & $\pm t_{\pm\mu\nu}$ &$-(t_{\pm}^{\mu\nu})^T$  & $\pm t_{\pm}^{\mu\nu}$  \\
 $\nabla^\mu\theta$ & $-\nabla^\mu\theta$ & $\nabla^\mu\theta$ & $-\nabla^\mu\theta$ \\
 $X$ & $-X$ & $X$ & $-X$\\
 \hline\hline
 \end{tabular}
 \end{table*}

 \section{From Building Blocks to Chiral Lagrangian}\label{lico}
 Given the properties of the building blocks in Table \ref{blbt}, we can construct the chiral Lagrangian by
 permuting and combining these building blocks. We shall focus mainly on the full $U(n_f)\times U(n_f)$ group,
 including all external sources and $\theta$ parameter ($v^\mu,a^\mu,s,p,\bar{t}^{\mu\nu},\theta$).
 Nevertheless, for orders above $O(p^4)$, the numbers of terms are very large.
 The higher the order, the much larger the number of terms becomes. Hence listing all terms is tedious.
 However, these terms are not linearly independent,
 we only need to find all linear conditions and list only those independent terms.
 All linear conditions come from the following relations.

 \begin{enumerate}
 \item Partial integration: Except for a total derivative term,
 we can remove one covariant derivative with respect to another operator in the following manner
 \begin{eqnarray}
 &&\la \nabla^\mu A B\cdots\ra\la CD\cdots\ra\cdots\notag\\
 &=&-\la A \nabla^\mu B\cdots\ra\la CD\cdots\ra\cdots-\cdots
 -\la A B\cdots\ra\la \nabla^\mu CD\cdots\ra\cdots -\la A B\cdots\ra\la C\nabla^\mu D\cdots\ra\cdots-\cdots,\label{pir}
 \end{eqnarray}
 where ``$\cdots$" represent one or more operators.

 \item Equations of motion: From the chiral Lagrangian and Euler-Lagrange equation,
 the lowest-order equations of motion(EOM) in the LR-basis for the chiral Lagrangian are given in Eq.(22) of Ref. \cite{U3}.
 In the building-blocks basis, we have
 \begin{eqnarray}
 \nabla_\mu u^\mu&=&\frac{i}{2}\frac{W'_0}{W_1}-\frac{i}{2}\frac{W'_1}{W_1}\la u^\mu u_\mu\ra
 +i\frac{W'_1}{W_1}(\la u_\mu\ra+i\nabla_\mu\thetah)u^\mu
 +\frac{i}{2}\frac{W_2}{W_1}\chim-\frac{1}{2}\frac{W_3}{W_1}\chip\notag\\
 &&-\frac{i}{2}\frac{W'_2}{W_1}\la\chip\ra+\frac{1}{2}\frac{W'_3}{W_1}\la\chim\ra
 +\bigg(\frac{i}{2}\frac{W'_5}{W_1}-\frac{i}{2}\frac{W'_6}{W_1}\bigg)(\nabla_\mu\thetah)(\nabla^\mu\thetah)
 +\frac{i}{2}\frac{W_5}{W_1}\nabla_\mu \nabla^\mu\thetah.\label{eomb}
 \end{eqnarray}
 The $W$ coefficients will be defined in (\ref{p2r}).
 Some minus signs in \eqref{eomb} are required to match the coefficients in \cite{U3}.
 The conclusion of \cite{p62,eom} was that up to $O(p^6)$, the lowest-order EOM are satisfied.
 If we want to evaluate a higher-order chiral Lagrangian, we need to develop the EOM to a higher order,
 which only adds terms on the r.h.s. of \eqref{eomb}.
 If we are only wanting the independent terms, we can also eliminate the factor $\nabla_\mu u^\mu$ without any difficulties.

 \item Bianchi identity: From Eqs.(\ref{Gam1}) , we can get
 \begin{eqnarray}
 \nabla^\mu\Gamma^{\nu\lambda}+\nabla^\nu\Gamma^{\lambda\mu}+\nabla^\lambda\Gamma^{\mu\nu}=0,\label{bi}
 \end{eqnarray}
 which gives a relation between the covariant derivatives of $\Gamma^{\mu\nu}(\fp^{\mu\nu})$.

 \item Schouten identity: When constructing the odd parity terms,
 we need to use the odd-intrinsic-parity factor $\epsilon^{\mu\nu\lambda\rho}$,
 as for example in the Schouten identity,
 \begin{eqnarray}
 \epsilon^{\mu\nu\lambda\rho}A^\sigma-\epsilon^{\sigma\nu\lambda\rho}A^\mu-\epsilon^{\mu\sigma\lambda\rho}A^\nu
 -\epsilon^{\mu\nu\sigma\rho}A^\lambda-\epsilon^{\mu\nu\lambda\sigma}A^\rho=0.\label{sir}
 \end{eqnarray}

 \item Tensor relations: For tensor sources, using (\ref{bas}) and (\ref{df}),
 one can obtain two relations between the odd-intrinsic-parity tensor $t_{-\mu\nu}$
 and the even-intrinsic-parity tensor $t_{+\mu\nu}$ \cite{tensor1};
 \begin{eqnarray}
 \epsilon_{\mu\nu\lambda\rho}t_{\pm}^{\lambda\rho}=2it_{\mp\mu\nu}.\label{tenr}
 \end{eqnarray}
 With these relations we can obtain the following transformations
 \begin{eqnarray}
 O_1\tp O_2\tm O_3 \leftrightarrow O_1\tm O_2\tp O_3,\text{ etc,}\label{tsr}
 \end{eqnarray}
 and develop relations between the two $t_{\pm}^{\mu\nu}$'s.

 \item Contact terms: In the chiral Lagrangian, we usually focus on the mesons fields,
 and separate the pure external sources terms and $\theta$ terms, called contact terms, leaving only meson dependent terms.

 \item Cayley-Hamilton relations: Usually, one develops the chiral Lagrangian for
 mesons composed of the lightest two or three quarks. In these situations,
 the operators are $2\times2$ or $3\times3$ matrices. The Cayley-Hamilton theorem states that
 every $n\times n$ matrix $A$ satisfies its own characteristic equation, $p(\lambda)$,
 \begin{eqnarray}
 &&p(\lambda)=\det(A-\lambda I_n)=0,\hspace{0.5cm} p(A)=0,\label{chr}
 \end{eqnarray}
 and hence gives a relation between the trace of an operator and its determinant.
 For $2\times2$ and $3\times3$ matrices,
 \begin{eqnarray}
 2\times2&:&\det(A)=\frac{1}{2}(\la A\ra^2-\la A^2\ra),\notag\\
 3\times3&:&\det(A)=\frac{1}{6}(\la A\ra^3-3\la A\ra\la A^2\ra+2\la A^3\ra).\label{matr}
 \end{eqnarray}
 Using \eqref{matr}, with some tricks (see \cite{p61,tensor1}),
 for arbitrary $2\times2$ matrices $A$ and $B$, we can get
 \begin{eqnarray}
 AB+BA-A\la B\ra-B\la A\ra-\la AB\ra+\la A\ra\la B\ra=0.\label{ch2}
 \end{eqnarray}
 For arbitrary $3\times3$ matrices $A,B$ and $C$, we can get
 \begin{eqnarray}
 0&=&ABC+ACB+BAC+BCA+CAB+CBA-AB\la C\ra-AC\la B\ra-BA\la C\ra-BC\la A\ra-CA\la B\ra\notag\\
 &&-CB\la A\ra-A\la BC\ra-B\la AC\ra-C\la AB\ra-\la ABC\ra-\la ACB\ra+A\la B\ra\la C\ra+B\la A\ra\la C\ra+C\la A\ra\la B\ra\notag\\
 &&+\la A\ra\la BC\ra+\la B\ra\la AC\ra+\la C\ra\la AB\ra-\la A\ra\la B\ra\la C\ra.\label{ch3}
 \end{eqnarray}
 Hence, when developing the two- or three-flavor chiral Lagrangian, the Cayley-Hamilton relations can not be ignored.
 \end{enumerate}

 The above conditions lead to all linear relations for the chiral Lagrangian.
 No other linear relations exist.
 Hence, if we can use all conditions systematically, the chiral Lagrangian can be obtained.
 The next section introduce our method in detail.

 \section{Construction of The Chiral Lagrangian}\label{consr}
 In this section, we work through the above situations one by one to construct the chiral Lagrangian.

 \subsection{Power Counting}
 In this paper, we only adopt the conventional power counting rule given in \cite{GS1,GS2},
 expand the chiral Lagrangian in powers of momentum.
 We show the power counting scheme in Table \ref{powc}. Each covariant derivative counts as $O(p^1)$.
 For tensor fields, we use the power counting rule in \cite{tensor1}, and consider tensor fields as combined (axial-)vector fields.
 Hence the chiral Lagrangian has only even orders in the expansion.
 \begin{table*}[h]
 \caption{\label{powc}Power counting of each base.}
 \begin{tabular}{cccc}
 \hline\hline order & LR-basis & building blocks \\
 \hline $O(p^0)$ & $U,\thetah,X$  & $u,\thetah,X$ \\
 $O(p^1)$ & $a^\mu,v^\mu$  & $u^\mu$ \\
 $O(p^2)$ & $\chi,t^{\mu\nu}$  & $\chi_{\pm},f_{\pm}^{\mu\nu},t_{\pm}^{\mu\nu}$\\
 \hline\hline
 \end{tabular}
 \end{table*}

 In Table \ref{blbt} and \ref{powc}, one can see that, $X$ is of order $O(p^0)$,
 hence in the chiral Lagrangian it can exist in functional form. $X$ also has odd parity,
 so a Lagrangian without $X$ can have both even or odd parity terms.
 The general form of the chiral Lagrangian to any given order is
 \begin{eqnarray}
 \mathscr{L}^{2n}=\sum_{m}f_m(X)O_m,\label{cl}
 \end{eqnarray}
 where the $f_m(X)$ are any functions of $X$, and $O_m$ are operators up to order $O(p^{2n})$.
 To construct the chiral Lagrangian of given order is to find all $O_m$ up to that order.

 \subsection{Reduced Building Blocks}
 The building blocks in Table \ref{blbt} are not very suitable for constructing the chiral Lagrangian directly,
 because some building blocks, such as $\fp^{\mu\nu}$, couple to different sources.
 Although the calculation is a little complex, we can break it up into smaller steps.
 First, before the calculations,
 we reduce the building blocks to more basic ones, each containing a given number of external sources.
 We call these the reduced building blocks, which we use when searching linear relations.
 At the end of the calculation, we revert to the original building blocks.

 From (\ref{Gam1}), we can see $f_{+}^{\mu\nu}$ is related to $\Gamma^{\mu\nu}$, with some redundant $u^\mu$.
 If we have only the linear independent terms, the choice of $\Gamma^{\mu\nu}$ is much more convenient,
 because it is directly relevant in the Bianchi identity(\ref{bi}).
 Ignoring the redundant $u^\mu$ makes no sense when we are only wanting the independent terms.
 Their effects can only be compensated in contributions of $u_{\mu}$.
 For the same reason, for (\ref{chimu}), we choose $\nabla^\mu\chi_{\pm}$ instead of $\chi_{\pm}^\mu$.
 We call those building blocks in Table \ref{blbt} along with the substitutions
 \begin{eqnarray}
 \fp^{\mu\nu}\longleftrightarrow i\Gamma^{\mu\nu},\hspace{0.2cm}
 \chi_{\pm}^\mu\longleftrightarrow\nabla^\mu\chi_\pm\label{rdb}
 \end{eqnarray}
 {\it the reduced building blocks}.
 The P, C and hermiticity properties between the original building blocks and the reduced building blocks are the same.

 If we obtain the high-order chiral Lagrangian, then appearing with the reduced building blocks;
 will be their covariant derivatives. To order $O(p^{2n})$, with the partial integration relations,
 we need $\nabla\nabla\cdots\nabla O$ to order $O(p^{n})$ at most, where $O$ is a reduced building block.

 Furthermore, for convenience, when we are only determining the independent operators\eqref{Gam0} can be simplified to
 \begin{eqnarray}
 [\nabla_\mu,\nabla_\nu]O\to0,\label{rnn}
 \end{eqnarray}
 because the excess terms on the r.h.s. of \eqref{Gam0} can be constructed by $\Gamma^{\mu\nu}$ and $O$ with two less $\nabla$.

 \subsection{Permutation, Combination and Primary Screening}\label{pcps}
 Permuting all reduced building blocks and their covariant derivatives, and adding appropriate traces,
 we can get a complete but not linearly independent set of operators.
 Because of \eqref{tenr}, $\tm^{\mu\nu}$ can appear once at most.
 Nevertheless the number of permutations is very large. A preliminary screening can lead to a simpler calculation.

 Most operators can be reduced using the Einstein summation convention $A^\mu A_\mu=A^\nu A_\nu$,
 and trace relations $\la AB\ra=\la BA\ra$. Even though the two operators are a little complex,
 we cannot assess immediately whether they are equal or not. A simple way is to change all operators to a unique form,
 we call it the {\it{standard form}}. We can assess whether two terms are equal directly by comparing their standard forms.
 \footnote{This is like a website verifying passwords in non-case-sensitive form.
 Changing all characters to upper or lower-case and then assessing them seems much easier.}

 First, we separate all operators and their indices,
 and assign a number to each operator and each index. Table \ref{noo} presents an example. For convenience,
 we consider the brackets (``$\la$" and ``$\ra$") and covariant derivative ($\nabla$) as single operators.
 \begin{table*}[h]
 \caption{\label{noo}Example of operator and index numbering.
 No significance is given to the specific numbers.
 These can be big or small, or even personal preferences.}
 \begin{tabular}{rccccccccccccc}
 \hline\hline operator & $\la$ & $\ra$ & $\nabla$ & $u$ & $h$ & $\chi_{+}$ & $\chi_{-}$ & $f_-$ & $\Gamma_+$
 & $t_{+}$ & $t_{-}$ & $\thetah$\\
 \hline number & 82 & 83 & 1004 & 1011 & 1058 & 1021 & 1141 & 1071 & 1007 & 2604 & 2605 & 1201  \\
 \hline index & $\mu$ & $\nu$ & $\lambda$ & $\rho$ & $\sigma$ \\
 \hline number & 1 & 2 & 3 & 4 & 5   \\
 \hline\hline
 \end{tabular}
 \end{table*}

 Second, we write down each operator numbers in order, and then enumerate the indices. We obtain a row vector
 such as
 \begin{eqnarray}
 \la u^\mu u^\nu h_{\mu\nu}\ra\to(82,1011,1011,1058,83,1,2,1,2).
 \end{eqnarray}
 We call it the {\it{numerical representation}} of an operator.
 With trace relations, Einstein summation convention and (anti-)symmetry index relations,
 we transform the numerical representation to its smallest possible form\footnote{We
 introduce a small number on the l.h.s. of the numerical representation to denote that it is the smallest form possible.}.
 The smallest representation is the {\it numerical representation},
 and the corresponding operator is the original operator in standard form.

 With the above procedure, all trivial relations are used, including the Einstein summation convention, trace relations and
 (anti-)symmetry indices relations.
 Next, we only need to work out one by one the conditions in Sec. \ref{lico}, with the standard form operators.
 For convenience, all the operators in the following calculations are in standard forms.

 \subsection{Substitutions and Classifications}
 Because $h^{\mu\nu}$ and $f_{\pm}^{\mu\nu}$ contain covariant derivatives,
 determine the linear relations is hard with the reduced building blocks.
 Using \eqref{df}, \eqref{fm} and \eqref{Gam1}, we can clearly reveal the covariant derivatives.

 In Sec.\ref{lico}, except for the EOM, each linear relation maintains its number of types of external sources.
 The relations cannot change scalar sources into vector sources or tensors to pseudoscalars for example\footnote{A
 tensor relation changes $\tp$ and $\tm$ by couples,
 so the total numbers of types of external sources remain the same.
 We also consider \eqref{rnn}.}.
 As the EOM is related to $\nabla^\mu u_\mu$, we only need to ignore terms including this factor.
 To simplify the problem, we classify all operator by the numbers of types of external sources.
 One classification contains the same numbers of each external sources.
 If there exists a linear relation, it only affects the operators in the same classification.
 Moreover, if we ignore the Cayley-Hamilton relations,
 or in the $n_f$-flavor case, the other relations can not change the trace number\footnote{The EOM
 has been used to exclude the $\nabla^\mu u_\mu$ factor}.

 Now we have classified the operator as a big set $C_{ij}$.
 Index $i$ indicates the sequence number of the types of the external sources,
 and $j$ indicates the number of traces. For any given pair $(i,j)$,
 all permutations of the reduced-building- block standard forms $D_{ij,k}$ are known,
 $k$ indicating the permutation numbers of the $C_{ij}$;
 all permutations of the revealable covariant derivatives standard forms $E_{ij,l}$ can be calculated,
 $l$ indicating the permutation number. Their linear relations are
 \begin{eqnarray}
 D_{ij,k}=\sum_{l}A_{ij,kl}E_{ij,l}.\label{lr0}
 \end{eqnarray}
 where $A_{ij,kl}$ are elements of their relative matrices.
 Our purpose is to find all the independent operators of $D_{ij,k}$,
 and $E_{ij,l}$ are the bridges in finding all linear relations.

 \subsection{Linear Relations}
 Generally, for the $n_f$-flavor case, the linear relations depend only on the small sets $D_{ij,k}$ or $E_{ij,l}$,
 our search scope is narrowed. All linear conditions can be added one by one
 \begin{enumerate}
 \item EOM: With the EOM \eqref{eomb}, we can remove all terms including ${h^\mu}_\mu$ in $D_{ij,k}$ or $\nabla^\mu u_\mu$ in $E_{ij,l}$.

 \item Partial integration, Bianchi identity, Schouten identity and tensor relations:
 The schemes for these conditions are the same. We can directly remove one term in \eqref{pir},\eqref{bi},
 \eqref{sir} and \eqref{tsr}. Although the relations are too many,
 remembering which relations have been used or not is not easy,
 and one does not know which $D_{ij,k}$ or $E_{ij,l}$ should be removed.
 A crude method is to apply these relations to all operators in $E_{ij,l}$.
 For partial integration, we only need to deal with the far left covariant derivative in each factor,
 because $E_{ij,l}$ is a complete set. There indeed exist some operators with another covariant derivative order.
 For the Bianchi identity, to reveal the covariant derivatives, \eqref{bi} needs to be changed to
 \begin{eqnarray}
 0&=&\nabla^{\mu}\nabla^{\nu}\Gam^{\lambda}-\nabla^{\mu}\nabla^{\lambda}\Gam^{\nu}
 -\nabla^{\mu}\Gam^{\nu}\Gam^{\lambda}-\Gam^{\nu}\nabla^{\mu}\Gam^{\lambda}
 +\nabla^{\mu}\Gam^{\lambda}\Gam^{\nu}+\Gam^{\lambda}\nabla^{\mu}\Gam^{\nu}\notag\\
 &&+\nabla^{\lambda}\nabla^{\mu}\Gam^{\nu}-\nabla^{\lambda}\nabla^{\nu}\Gam^{\mu}
 -\nabla^{\lambda}\Gam^{\mu}\Gam^{\nu}-\Gam^{\mu}\nabla^{\lambda}\Gam^{\nu}
 +\nabla^{\lambda}\Gam^{\nu}\Gam^{\mu}+\Gam^{\nu}\nabla^{\lambda}\Gam^{\mu}\notag\\
 &&+\nabla^{\nu}\nabla^{\lambda}\Gam^{\mu}-\nabla^{\nu}\nabla^{\mu}\Gam^{\lambda}
 -\nabla^{\nu}\Gam^{\lambda}\Gam^{\mu}-\Gam^{\lambda}\nabla^{\nu}\Gam^{\mu}
 +\nabla^{\nu}\Gam^{\mu}\Gam^{\lambda}+\Gam^{\mu}\nabla^{\nu}\Gam^{\lambda}.
 \end{eqnarray}
 We also only need to focus on the covariant derivative factor near a $\Gamma$.
 The other covariant derivatives, even if the first in $\nabla\nabla\Gamma$, need not be considered.
 For Schouten identity, we must exchange all indices different from that of $\epsilon$.
 For the tensor relations, tensors involve one $\tm$ and several $\tp$ terms.
 Because the number of $\tm$ is at most one, exchanging it for each $\tp$ can give a linear relations.
 In summary, we can write down all linear relations.
 \begin{eqnarray}
 \sum_l R_{ij,rl}E_{ij,l}=0,\label{cr0}
 \end{eqnarray}
 where $r$ is the number of relations, and $R_{ij,rl}$ are the relation matrices with the row index ``$ij$"
 and column index ``$rl$".
 The reduced row echelon form of $R_{ij,rl}$,
 \begin{eqnarray}
 R_{ij,rl}\to S_{ij,rl}=
 \left(\begin{array}{cccccccccc}
 1 & O & O & O & O & \cdots & \cdots & \cdots \\
   &   & 1 & O & O & \cdots & \cdots & \cdots \\
   &   &   & \cdots & \cdots & \cdots & \cdots & \cdots \\
   &   &   &   & 1 & \cdots & \cdots & \cdots \\
   &   &   &   &   &   O    &    O   & O
 \end{array}\right),\label{lrs0}
 \end{eqnarray}
 where the $O$ are some appropriate dimension-zero-matrices and ``$\cdots$" can be non-zero matrices.
 The rank of $R_{ij,rl}$ or $S_{ij,rl}$ is equal to the number of linear relations.
 And each non-zero row-vector in $S_{ij,rl}$ gives a linear relation.
 We consider numbers of the operators that have number ``1" as independent.
 Hence (\ref{lr0}) can be reduced to
 \begin{eqnarray}
 D_{ij,k}=\sum_{l}A'_{ij,kl}E_{ij,l},\label{lr1}
 \end{eqnarray}
 where using \eqref{cr0} $A'_{ij,kl}$ are the matrices $A_{ij,kl}$,
 with all linear dependent operators removed in the columns of \eqref{lrs0} with ``1".

 \item Cayley-Hamilton relations: Because the Cayley-Hamilton relations can change the trace,
 we collect all $D_{ij}$ and $E_{ij}$, as
 \begin{eqnarray}
 &&D_{ij}\to D_{i}=[D_{i1},D_{i2},D_{i3},\cdots,D_{in}]\notag\\
 &&E_{ij}\to E_{i}=[E_{i1},E_{i2},E_{i3},\cdots,E_{in}],\label{ij2i}
 \end{eqnarray}
 where $n$ is the number corresponding to the maximum trace. Then, (\ref{lr1}) is reduced to
 \begin{eqnarray}
 D_{i,k}=\sum_{l}A'_{i,kl}E_{i,l};\label{lr2}
 \end{eqnarray}
 the Cayley-Hamilton relations only give new linear relations when one trace contain more than $n_f$ operators.
 $A,B,C$ in \eqref{ch2} and \eqref{ch3} can be one factor or more.
 Applying all possible combinations, we can get all Cayley-Hamilton relations.
 Removing all linear dependent terms, as in \eqref{lr1}, we obtain
 \begin{eqnarray}
 D_{n_f,i,k}=\sum_{l}A''_{n_f,i,kl}E_{n_f,i,l},\label{lr3}
 \end{eqnarray}
 where $D_{n_f,i,k}$ is the $k$th operator in the $n_f$-flavor case in the classification $i$ ($n_f=2$ or $3$),
 $E_{n_f,i,l}$ is the $l$th revealable covariant derivative operator in the $n_f$-flavor case in the classification $i$,
 and $A''_{n_f,i,kl}$ is the matrix of coefficients relating the two.

 \item Contact terms: Contact terms can be found in a specialized treatment.
 Because all the building blocks in Table \ref{blbt} mix $u$ (or $U$) fields and external sources,
 Constructing the contact terms with such building blocks is not so useful.
 Hence we construct contact terms with the LR-basis, without $U$ fields.
 Their relations are
 \begin{eqnarray}
 F_L^{\mu\nu}&=&\frac{1}{2}u^\dag(f_+^{\mu\nu}+f_-^{\mu\nu})u,\notag\\
 F_R^{\mu\nu}&=&\frac{1}{2}u(f_+^{\mu\nu}-f_-^{\mu\nu})u^\dag,\notag\\
 \chi&=&\frac{1}{2}u(\chi_++\chi_-)u,\notag\\
 \chi^\dag&=&\frac{1}{2}u^\dag(\chi_+-\chi_-)u^\dag\notag\\
 t^{\mu\nu}&=&\frac{1}{2}\xu(\tp^{\mu\nu}+\tm^{\mu\nu})\xu,\notag\\
 \td^{\mu\nu}&=&\frac{1}{2}\ud(\tp^{\mu\nu}-\tm^{\mu\nu})\ud\notag\\
 \nabla^\mu\thetah&=&\nabla^\mu\thetah.\label{lrct}
 \end{eqnarray}
 To construct the contact terms, one only needs the basis and their covariant derivatives in \eqref{lrct}.
 By performing the same steps above, we can obtain all permutations of the non-$U$ operators.
 There exist though two exceptions.
 The first is that different LR-bases have different chiral rotations.
 Hence, not all permutations, but only the chiral invariant operators remain.
 The second is that for SU group, $u(U)$ is unitary, so $\det u(\det U)=1$.
 With \eqref{matr}, the combination of the determinant of the LR-basis or its covariant derivatives can be chiral invariant as well.
 At order $p^2$, there exists no such operator. For order $p^4$, there exist one term only when $n_f=2$,
 \begin{eqnarray}
 &&\det \chi+\det\chi^\dag
 =-\frac{1}{4}\la\chip^2\ra-\frac{1}{4}\la\chim^2\ra+\frac{1}{4}\la\chip\ra\la\chip\ra+\frac{1}{4}\la\chim\ra\la\chim\ra.\label{ct42}
 \end{eqnarray}
 For order $p^6$, there exist one term when $n_f=3$,
 \begin{eqnarray}
 \det \chi+\det\chi^\dag
 &=&\frac{1}{24}\la\chip\ra^3+\frac{1}{8}\la\chip\ra\la\chim\ra^2
 -\frac{1}{8}\la\chip\ra\la\chip^2\ra-\frac{1}{8}\la\chip\ra\la\chim^2\ra\notag\\
 &&-\frac{1}{4}\la\chip\chim\ra\la\chim\ra
 +\frac{1}{12}\la\chip^3\ra+\frac{1}{4}\la\chip\chim^2\ra.\label{ct63}
 \end{eqnarray}
 When $n_f=2$, there exist three terms,
 \begin{eqnarray}
 \det D^{\mu}\chi+\det D^{\mu}\chi^\dag&=&-\frac{1}{4}\la\chip^{\mu}\chi_{+\mu}\ra-\frac{1}{4}\la\chim^{\mu}\chi_{-\mu}\ra
 +\frac{1}{4}\la\chip^{\mu}\ra\la\chi_{+\mu}\ra+\frac{1}{4}\la\chim^{\mu}\ra\la\chi_{-\mu}\ra\notag\\
 (\detx+\detxd)\nabla^{\mu}\thetah\nabla_{\mu}\thetah
 &=&\Big(-\frac{1}{4}\la\chip^2\ra-\frac{1}{4}\la\chim^2\ra
 +\frac{1}{4}\la\chip\ra\la\chip\ra+\frac{1}{4}\la\chim\ra\la\chim\ra\Big)\nabla^{\mu}\thetah\nabla_{\mu}\thetah\notag\\
 (\detx+\detxd)\nabla^{\mu}\nabla_{\mu}\thetah
 &=&\Big(-\frac{1}{4}\la\chip^2\ra-\frac{1}{4}\la\chim^2\ra
 +\frac{1}{4}\la\chip\ra\la\chip\ra+\frac{1}{4}\la\chim\ra\la\chim\ra\Big)\nabla^{\mu}\nabla_{\mu}\thetah.\label{ct62}
 \end{eqnarray}
 The first term is similar to $\la D_\mu\chi D^\mu\tilde\chi\ra+$h.c. given in \cite{p6a2}.

 Following the steps above, we obtain a similar relation as in \eqref{lr1} or \eqref{lr2}.
 \begin{eqnarray}
 D_{n_f,k}=\sum_{l}A'_{n_f,kl}E_{n_f,l},\hspace{0.2cm}n_f=3\text{ or }2.\label{lrcr}
 \end{eqnarray}
 $D_{n_f,k}$ and $E_{n_f,k}$ are all $D_{ij}$ and $E_{ij}$ operators in \eqref{ij2i} with respective contact terms.

 \item Eliminating rule: For a given $n_f$, collecting all the linear relations above, we can obtain
 \begin{eqnarray}
 \sum_l R_{rl}D_{l}=0\label{cr1}
 \end{eqnarray}
 The linear independent rows of $R_{rl}$ correspond to independent terms.
 The question arises: Which terms are considered linear dependent and need to be removed?
 We follow the rule: The contact terms and terms with the least number of covariant derivatives
 and traces are retained as far as possible.
 Furthermore, we retain the smallest-possible numerical representation of the operator (see Sec. \ref{pcps}).

 \item Real: The chiral Lagrangian is real. Finally,
 with the last column in Table \ref{lrbhb} and \ref{blbt}, some terms need factoring by ``$i$" to make them real.

 \item U and SU groups: The above discussions are about U group. Most of the time, one does not care about the $\eta'$ particle,
 and hence the discussion is limited in SU group. Of course, one can integrate $\eta'$ to match the SU group results \cite{U3p},
 but to do so gives nothing of interest.
 It is convenient when identifying the terms involving SU group in the U group.
 However under the SU group, all the operators $O_m$ in \eqref{cl} are even parity.
 Usually, we focus on traceless sources ($\la v^\mu\ra=\la a^\mu\ra=0$ or $\la u^{\mu}\ra=\la \fp^{\mu\nu}\ra=0$) in the SU group.
 although sometimes we need study to the singlet (axial-)vector external sources, so $\la u^{\mu}\ra\neq0$ or $\la \fp^{\mu\nu}\ra\neq0$
 need to be considered. Fortunately, these situations cannot change the calculations above;
 we only need to set traceless terms to zero.
 \end{enumerate}

 \subsection{Discussions}
 \begin{enumerate}
 \item Linear relations of the dependent operators:
   At this point, the chiral Lagrangian can be obtained,
   but when we solve other problems, we may meet some dependent operators.
   Most of their expansions in terms of independent operators are simple, but several are complex;
   see Appendix B in \cite{p62} and \cite{p6p} for examples.
   We will introduce a scheme to generate these relations.
   When we derive the chiral Lagrangian, all the operator are $O_n$, and all revealable
   covariant derivative standard forms are $P_m$. We can get their linear relations from \eqref{lr1},
   and full relations from \eqref{Gam0},\eqref{Gam1} and \eqref{chimu};
   \begin{eqnarray}
   O_n=\sum_m B_{nm}P_m
   \end{eqnarray}
   where $B_{nm}$ is the matrix of coefficients, which have been used in all linear relations.
   If $O'$ is another operator, then
   \begin{eqnarray}
   O'=\sum_{{m}}T_{m}P_m.\label{opr}
   \end{eqnarray}
   Using all linear relations, this can reduced to
   \begin{eqnarray}
   O'=\sum_{{m}}T'_{m}P_m.\label{oprp}
   \end{eqnarray}
   Supposing the relation between $O'$ and the set of $O_n$ is
   \begin{eqnarray}
   O'=\sum_{n}C_n O_n.
   \end{eqnarray}
   The coefficients $C_n$ are the solution of the linear equation,
   \begin{eqnarray}
   \sum_n B_{nm}C_n=T'_{m}
   \end{eqnarray}
   Hence all linear relations between linear dependent terms and linear independent terms can be obtained
   such as in \cite{p6p} and Appendix B in \cite{p62}.

   \item  The steps above can be done directly, and systematically, without any artifices,
   and hence can be performed on a computer. Furthermore, the accuracy is reliable,
   and we can obtain the results relatively quickly.
   This method is not very sensitive with regard to the form of the linear relations;
   we can add or remove some linear relation in \eqref{cr0},
   without affecting other steps.
   Therefore it can be used for other problems where Lagrangian need to be constructed,
   for instance, the electroweak chiral Lagrangian \cite{ew1,ew2}.
   If we solve some special problems, changing the counting rules can be more effective \cite{zhanghh}.
 \end{enumerate}

 \section{Results}\label{res}
 We can now obtain the chiral Lagrangian to order $p^6$. We list the results in the following.
 \subsection{$p^0$ and $p^2$ order}
 The results for order $p^0$ and $p^2$ are the same as in \cite{U3},
 \begin{eqnarray}
 \mathscr{L}_{0+2}&=&-W_0(X)+W_1(X)\la u^2\ra+W_2(X)\la\chip\ra+iW_3(X)\la\chim\ra
 -W_4(X)\la u^{\mu}\ra\la u_{\mu}\ra\notag\\
 &&-iW_5(X)\la u^{\mu}\ra\nabla^\mu\thetah+W_6(X)\nabla^\mu\thetah\nabla_\mu\thetah.\label{p2r}
 \end{eqnarray}
 Minus signs of some of the coefficients are to confirm that in \cite{U3}.

 \subsection{$p^4$ order}\label{p4result}
 For the U groups, results have been given in \cite{U3}. For the SU groups, results can be found in \cite{GS1,GS2},
 or \cite{p6r,p6rn} in terms of the building blocks, or \cite{tensor1} with tensor sources.
 We have repeated the calculations to check our method.
 We find one more linear relations than \cite{U3} for the U group.
 \begin{eqnarray}
 i\epsilon_{\mu\nu\lambda\rho}\la(F_L^{\mu\nu}+\Ud F_R^{\mu\nu}U)C^{\lambda}C^{\rho}\ra
 =\epsilon_{\mu\nu\lambda\rho}\la F_L^{\mu\nu}\Ud F_R^{\lambda\rho}U\ra,\text{ or }
 O_{29}=O_{28}.\label{p4more}
 \end{eqnarray}
 The meanings of the notations in \eqref{p4more} can be found in \cite{U3}.
 It can be proved by substituting \eqref{flr} and $C^\mu=U^\dag D^\mu U$ into \eqref{p4more} directly.
 By transforming \eqref{p4more} to the building-block basis, the calculation is much simpler.
 We list our results in Table \ref{p4rtab} in Appendix \ref{p4rsec}.
 Besides \eqref{p4more}, there are some difference between Table \ref{p4rtab} and \cite{U3}:
 Aside from the U(N) and U(3) group, we also include the U(2) groups results
 and the tensor sources as well. For the SU groups, we considered the following three types of external sources,
 \begin{eqnarray}
 \begin{array}{lll}
 \text{SU(3)}_\text{I}&:&\la\fp^{\mu\nu}\ra\neq0,\la u^{\mu}\ra=0\\
 \text{SU(3)}_\text{II}&:&\la\fp^{\mu\nu}\ra=0,\la u^{\mu}\ra\neq0\\
 \text{SU(3)}_\text{III}&:&\la\fp^{\mu\nu}\ra=0,\la u^{\mu}\ra=0
 \end{array}
 \end{eqnarray}
 The results with $\la\fp^{\mu\nu}\ra\neq0$ and $\la u^{\mu}\ra\neq0$ can be found within the U group results,
 so are not listed here.
 The SU(3)$_\text{III}$ case is the conventional SU group case.

 \subsection{$p^6$ order}\label{p6result}
 At order $p^6$, as much a check of our method as of existing results,
 we also compare our results with that of others.
 We recover the results given in \cite{p62,p6p,p6a2} without tensor sources. With tensor sources,
 apart from the more relations in \cite{ourt} than \cite{tensor1},
 we find one other linear relation in the $n_f$-flavor case,
 \begin{eqnarray}
 Y_{102}=\frac{1}{2}Y_{78}-Y_{87}-2Y_{101}-Y_{117},
 \end{eqnarray}
 and another nine linear relations in the two-flavor case,
 \begin{eqnarray}
 Y_{48}&=&-Y_{47}\notag\\
 Y_{49}&=&-Y_{47}\notag\\
 Y_{50}&=&2Y_{47}\notag\\
 Y_{71}&=&-Y_{70}\notag\\
 Y_{72}&=&-Y_{70}\notag\\
 Y_{73}&=&-2Y_{70}\notag\\
 Y_{92}&=&0\notag\\
 Y_{106}&=&\frac{1}{2}Y_{9}+\frac{1}{2}Y_{10}-2Y_{11}-Y_{13}+Y_{51}-4Y_{52}+Y_{82}+2Y_{90}-2Y_{105}+8Y_{118}\notag\\
 Y_{108}&=&-\frac{1}{8}Y_{9}-\frac{1}{8}Y_{10}+\frac{1}{2}Y_{11}+\frac{1}{4}Y_{13}+\frac{1}{8}Y_{32}
 -\frac{1}{16}Y_{35}-\frac{1}{4}Y_{51}\notag\\
 &&+Y_{52}-\frac{1}{2}Y_{82}-\frac{1}{2}Y_{90}+Y_{105}-2Y_{118}
 \end{eqnarray}
 At order $p^6$, the number of independent terms is too large, so we separate the chiral Lagrangian into four types,
 depending on whether they include $t^{\mu\nu}$ or $\thetah$.
 We list these four types in the first column of Table \ref{np6}, and summarize the number of terms in each type.
 Explicit terms can be found in Section \ref{p6rsec}.
 Table \ref{p6tab1} and \ref{p6tab2} give the even- and odd-parity results when $\theta=\tb^{\mu\nu}=0$ respectively;
 similarly for
 Table \ref{p6tab3} and \ref{p6tab4} when $\theta\neq0,\tb^{\mu\nu}=0$,
 Table \ref{p6tab5} and \ref{p6tab6} when $\theta=0,\tb^{\mu\nu}\neq0$ respectively,
 and Table \ref{p6tab7} and \ref{p6tab8} when $\theta\neq0,\tb^{\mu\nu}\neq0$.
 These are some differences in sequence numbers from \cite{p62,p6a2,tensor1}, but this creates no ambiguity.
 \begin{table*}[h]
 \caption{\label{np6}Numbers of terms in the chiral Lagrangian at order $p^6$.}
 \begin{tabular}{l|c|@{\tiny}c|c|c|c|c|c|c|c|c|c|c|c}
 \hline\hline Number & $P=+/-$ & \multicolumn{3}{c|}{U(N)}
 & \multicolumn{3}{c|}{SU(N)$_\mathrm{I}$}
 & \multicolumn{3}{c|}{SU(N)$_\mathrm{II}$}
 & \multicolumn{3}{c}{SU(N)$_\mathrm{III}$} \\
 \hline classification & $P$ & $n$ & $3$ & $2$ & $n$ & $3$ & $2$ & $n$ & $3$ & $2$ & $n$ & $3$ & $2$\\
 \hline$\theta=\tb^{\mu\nu}=0$ & $287/206$ & 493 & 451 & 281 & 155 & 133 & 74 & 263 & 235 & 142 & 138 & 116 & 60 \\
 $\theta\neq0,\tb^{\mu\nu}=0$ & $261/206$ & 467 & 455 & 371 & 124 & 118 & 83 & 251 & 243 & 194 & 116 & 110 & 75 \\
 $\theta=0,\tb^{\mu\nu}\neq0$ & $159/149$ & 308 & 297 & 209 & 110 & 104 & 67 & 137 & 131 & 88 & 97 & 91 & 56 \\
 $\theta\neq0,\tb^{\mu\nu}\neq0$ & $64/59$ & 123 & 123 & 108 & 51 & 51 & 43 & 52 & 52 & 44 & 42 & 42 & 34 \\
 total & $771/620$ & 1391 & 1326 & 969 & 440 & 406 & 267 & 703 & 661 & 468 & 393 & 359 & 225 \\
 \hline\hline
 \end{tabular}
 \end{table*}

 \section{Summary}\label{summ}

 In this paper, we constructed the full $p^6$-order U group chiral Lagrangian,
 including $n_f$-, three- and two-flavor cases, all bilinear currents, and the $\theta$ parameter,
 and three kinds of SU groups results.
 The number of terms in each case can be found in Table \ref{np6},
 and the detailed form can be found in Appendix \ref{p6rsec}.
 We also list the full $p^4$-order results in Appendix \ref{p4rsec}.
 During the calculations, we found one extra linear relation in the $p^4$-order under the U(3) group,
 and one more linear relation in the $p^6$-order under the SU groups with tensor sources in $n_f$ flavors,
 and further nine more relations in the two-flavor case.
 We develop a scheme to derive the relations among the linear dependent and independent terms
 which enable us to systematically set and examine various possible relations among the LECs.

 \section*{Acknowledgments}
 This work was supported by the National Science Foundation of China
 (NSFC) under Grants No. 11205034 and No.11075085;  National Basic Research Program of China (973
Program) under Grants No. 2010CB833000; the Specialized Research Fund for the Doctoral Program of High Education of China No. 20110002110010 and France China Particle Physics Laboratory (FCPPL) financial support; the Natural Science Foundation of Guangxi Grants No. 2013GXNSFBA019012;
 and Special Foundation for Distinguished Expert Program of Guangxi; the Tsinghua University Initiative Scientific Research Program;
 and the Scientific Research Foundation of GuangXi University Grant No. XBZ100686.

 \appendix
 \section{The $p^4$ order results}\label{p4rsec}
\begin{longtable}{l|c|c|c|c|c|c|c|c|c|c|c|c|c}
 \caption{\label{p4rtab}The $p^4$ order results in U groups and SU groups.
 The numbers are the sequence numbers in each case.
 The details can be found in Section \ref{p4result}}\\
 \hline\hline &  & \multicolumn{3}{|c|}{U(3)}
 & \multicolumn{3}{|c|}{SU(3)$_\mathrm{I}$}
 & \multicolumn{3}{|c|}{SU(3)$_\mathrm{II}$}
 & \multicolumn{3}{|c}{SU(3)$_\mathrm{III}$} \\
 \hline operators & $P$ & $n$ & $3$ & $2$ & $n$ & $3$ & $2$ & $n$ & $3$ & $2$ & $n$ & $3$ & $2$\\
 \hline\endfirsthead

 \hline\hline operators & $P$ & $n$ & $3$ & $2$ & $n$ & $3$ & $2$ & $n$ & $3$ & $2$ & $n$ & $3$ & $2$\\
 \hline\endhead

 \hline\hline 
 \endfoot

 \hline\endlastfoot

 $ \la u^{\mu}u_{\mu}u^{\nu}u_{\nu}\ra
 $ & $+$ & 1 & 1 & 1 & 1 & 1 & 1 & 1 & 1 & 1 & 1 & 1 & 1 \\
 $ \la u^{\mu}u^{\nu}u_{\mu}u_{\nu}\ra
 $ & $+$ & 2 & 2 & 2 & 2 & 2 & 2 & 2 & 2 & 2 & 2 & 2 & 2 \\
 $ \la u^{\mu}\ra\la u_{\mu}u^{\nu}u_{\nu}\ra
 $ & $+$ & 3 & 3 & 3 &  &  &  & 3 & 3 & 3 &  &  &  \\
 $ \la u^{\mu}u_{\mu}\ra\la u^{\nu}u_{\nu}\ra
 $ & $+$ & 4 & 4 & 4 & 3 & 3 &  & 4 & 4 & 4 & 3 & 3 &  \\
 $ \la u^{\mu}u^{\nu}\ra\la u_{\mu}u_{\nu}\ra
 $ & $+$ & 5 & 5 & 5 & 4 &  &  & 5 & 5 & 5 & 4 &  &  \\
 $ \la u^{\mu}\ra\la u_{\mu}\ra\la u^{\nu}u_{\nu}\ra
 $ & $+$ & 6 & 6 &  &  &  &  & 6 & 6 &  &  &  &  \\
 $ \la u^{\mu}\ra\la u^{\nu}\ra\la u_{\mu}u_{\nu}\ra
 $ & $+$ & 7 & 7 &  &  &  &  & 7 & 7 &  &  &  &  \\
 $ \la u^{\mu}\ra\la u_{\mu}\ra\la u^{\nu}\ra\la u_{\nu}\ra
 $ & $+$ & 8 &  &  &  &  &  & 8 &  &  &  &  &  \\
\hline\hline
 \end{longtable}

 \section{The $p^6$ order results}\label{p6rsec}
 {\tabcolsep 1pt\renewcommand{\arraystretch}{0.4}
 \begin{longtable}{l|c|c|c|c|c|c|c|c|c|c|c|c|l|c|c|c|c|c|c|c|c|c|c|c|c}
 \caption{\label{p6tab1}The $p^6$ order results in U groups and SU groups, with $t=0$, $\thetah=0$ and the even parity.
 The numbers are the sequence numbers in each case.
 The details can be found in Section \ref{p6result}.}\\

 \hline\hline & \multicolumn{3}{c|}{\zh U(3)}
 & \multicolumn{3}{c|}{\zh SU(3)$_\mathrm{I}$}
 & \multicolumn{3}{c|}{\zh SU(3)$_\mathrm{II}$}
 & \multicolumn{3}{c|}{\zh SU(3)$_\mathrm{III}$}
 & & \multicolumn{3}{c|}{\zh U(3)}
 & \multicolumn{3}{c|}{\zh SU(3)$_\mathrm{I}$}
 & \multicolumn{3}{c|}{\zh SU(3)$_\mathrm{II}$}
 & \multicolumn{3}{c}{\zh SU(3)$_\mathrm{III}$} \\

 \hline \zh operators & \zh $n$ & \zh $3$ & \zh $2$ & \zh $n$ & \zh $3$ & \zh $2$ & \zh $n$ & \zh $3$ & \zh $2$ & \zh $n$ & \zh $3$ & \zh $2$
 & \zh operators & \zh $n$ & \zh $3$ & \zh $2$ & \zh $n$ & \zh $3$ & \zh $2$ & \zh $n$ & \zh $3$ & \zh $2$ & \zh $n$ & \zh $3$ & \zh $2$\\
 \hline\endfirsthead

 \hline\hline \zh operators & \zh $n$ & \zh $3$ & \zh $2$ & \zh $n$ & \zh $3$ & \zh $2$ & \zh $n$ & \zh $3$ & \zh $2$ & \zh $n$ & \zh $3$ & \zh $2$
 & \zh operators & \zh $n$ & \zh $3$ & \zh $2$ & \zh $n$ & \zh $3$ & \zh $2$ & \zh $n$ & \zh $3$ & \zh $2$ & \zh $n$ & \zh $3$ & \zh $2$\\
 \hline\endhead

 \hline\hline 
 \endfoot

 \hline\endlastfoot

 \tiny$ \la u^{\mu}u_{\mu}u^{\nu}u_{\nu}u^{\lambda}u_{\lambda}\ra
 $ &\tiny 1 &\tiny 1 &\tiny 1 &\tiny 1 &\tiny 1 &\tiny 1 &\tiny 1 &\tiny 1 &\tiny 1 &\tiny 1 &\tiny 1 &\tiny 1
 & \tiny$ \la\chip f_{-}^{\mu\nu}f_{-\mu\nu}\ra
 $ &\tiny 146 &\tiny 123 &\tiny 72 &\tiny 73 &\tiny 55 &\tiny 30 &\tiny 135 &\tiny 112 &\tiny 64 &\tiny 68 &\tiny 50 &\tiny 28 \\
 \tiny$ \la u^{\mu}u_{\mu}u^{\nu}u^{\lambda}u_{\nu}u_{\lambda}\ra
 $ &\tiny 2 &\tiny 2 &\tiny 2 &\tiny 2 &\tiny 2 &\tiny 2 &\tiny 2 &\tiny 2 &\tiny 2 &\tiny 2 &\tiny 2 &\tiny 2
 & \tiny$ \la\chip h^{\mu\nu}h_{\mu\nu}\ra
 $ &\tiny 147 &\tiny 124 &\tiny 73 &\tiny 74 &\tiny 56 &\tiny 31 &\tiny 136 &\tiny 113 &\tiny 65 &\tiny 69 &\tiny 51 &\tiny 29 \\
 \tiny$ \la u^{\mu}u_{\mu}u^{\nu}u^{\lambda}u_{\lambda}u_{\nu}\ra
 $ &\tiny 3 &\tiny 3 &\tiny 3 &\tiny 3 &\tiny 3 &  &\tiny 3 &\tiny 3 &\tiny 3 &\tiny 3 &\tiny 3 &
 & \tiny$ \la u^{\mu}\chip\nabla^{\nu}f_{-\mu\nu}\ra
 +h.c.$ &\tiny 148 &\tiny 125 &\tiny 74 &\tiny 75 &\tiny 57 &\tiny 32 &\tiny 137 &\tiny 114 &\tiny 66 &\tiny 70 &\tiny 52 &\tiny 30 \\
 \tiny$ \la u^{\mu}u^{\nu}u_{\mu}u^{\lambda}u_{\nu}u_{\lambda}\ra
 $ &\tiny 4 &\tiny 4 &\tiny 4 &\tiny 4 &\tiny 4 &\tiny 3 &\tiny 4 &\tiny 4 &\tiny 4 &\tiny 4 &\tiny 4 &\tiny 3
 & \tiny$ \la\chip\ra\la f_{-}^{\mu\nu}f_{-\mu\nu}\ra
 $ &\tiny 149 &\tiny 126 &\tiny 75 &\tiny 76 &\tiny 58 &  &\tiny 138 &\tiny 115 &\tiny 67 &\tiny 71 &\tiny 53 &  \\
 \tiny$ \la u^{\mu}u^{\nu}u^{\lambda}u_{\mu}u_{\nu}u_{\lambda}\ra
 $ &\tiny 5 &\tiny 5 &\tiny 5 &\tiny 5 &\tiny 5 &  &\tiny 5 &\tiny 5 &\tiny 5 &\tiny 5 &\tiny 5 &
 & \tiny$ \la\chip f_{-}^{\mu\nu}\ra\la f_{-\mu\nu}\ra
 $ &\tiny 150 &\tiny 127 &\tiny 76 &  &  &  &\tiny 139 &\tiny 116 &\tiny 68 &  &  &  \\
\hline\hline
 \end{longtable}}

 {\tabcolsep 1pt\renewcommand{\arraystretch}{0.4}
 \renewcommand{\oddsidemargin}{-1cm}
 \begin{longtable}{@{\hspace{-1cm}}l|c|c|c|l|c|c|c|l|c|c|c|l|c|c|c}
 \caption{\label{p6tab2}The $p^6$ order results in U groups and SU groups, with $t=0$, $\thetah=0$ and the odd parity.
 The numbers are the sequence numbers in each case.
 The details can be found in Section \ref{p6result}.}\\

 \hline\hline & \multicolumn{3}{c|}{\zh U(3)}
 & & \multicolumn{3}{c|}{\zh U(3)}
 & & \multicolumn{3}{c|}{\zh U(3)}
 & & \multicolumn{3}{c}{\zh U(3)} \\
 \hline \zh operators & \zh $n$ & \zh $3$ & \zh $2$ &\zh operators & \zh $n$ & \zh $3$ & \zh $2$
 &\zh operators & \zh $n$ & \zh $3$ & \zh $2$ &\zh operators & \zh $n$ & \zh $3$ & \zh $2$ \\
 \hline\endfirsthead

 \hline\hline \zh operators & \zh $n$ & \zh $3$ & \zh $2$ &\zh operators & \zh $n$ & \zh $3$ & \zh $2$
 &\zh operators & \zh $n$ & \zh $3$ & \zh $2$ &\zh operators & \zh $n$ & \zh $3$ & \zh $2$ \\
 \hline\endhead

 \hline\hline 
 \endfoot

 \hline\endlastfoot

 \tiny$ \la u^{\mu}u_{\mu}u^{\nu}u^{\lambda}f_{-\nu\lambda}\ra
 +h.c.$ &\tiny 288 &\tiny 260 &\tiny 164
 & \tiny$ \la u^{\mu}\chip u^{\nu}h_{\mu\nu}\ra
 $ &\tiny 340 &\tiny 308 &\tiny 192
 & \tiny$ \la f_{+}^{\mu\nu}u_{\mu}u_{\nu}\chim\ra
 +h.c.$ &\tiny 392 &\tiny 355 &\tiny 221
 & \tiny$i \epsilon^{\mu\nu\lambda\rho}\la f_{+\mu\nu}u_{\lambda}u_{\rho}u^{\sigma}u_{\sigma}\ra
 +h.c.$ &\tiny 444 &\tiny 405 &\tiny 254 \\
 \tiny$ \la u^{\mu}u^{\nu}u_{\mu}u^{\lambda}f_{-\nu\lambda}\ra
 +h.c.$ &\tiny 289 &\tiny 261 &\tiny 165
 & \tiny$ \la u^{\mu}\ra\la u^{\nu}\chip f_{-\mu\nu}\ra
 +h.c.$ &\tiny 341 &\tiny 309 &\tiny 193
 & \tiny$ \la f_{+}^{\mu\nu}u_{\mu}\chim u_{\nu}\ra
 $ &\tiny 393 &\tiny 356 &\tiny 222
 & \tiny$i \epsilon^{\mu\nu\lambda\rho}\la f_{+\mu\nu}u_{\lambda}u^{\sigma}u_{\rho}u_{\sigma}\ra
 +h.c.$ &\tiny 445 &\tiny 406 &\tiny 255 \\
 \tiny$ \la u^{\mu}u_{\mu}u^{\nu}u^{\lambda}h_{\nu\lambda}\ra
 +h.c.$ &\tiny 290 &\tiny 262 &\tiny 166
 & \tiny$ \la u^{\mu}\chip\ra\la u^{\nu}f_{-\mu\nu}\ra
 $ &\tiny 342 &\tiny 310 &\tiny 194
 & \tiny$ \la f_{+}^{\mu\nu}\ra\la u_{\mu}u_{\nu}\chim\ra
 $ &\tiny 394 &\tiny 357 &\tiny 223
 & \tiny$i \epsilon^{\mu\nu\lambda\rho}\la f_{+\mu\nu}u_{\lambda}u^{\sigma}u_{\sigma}u_{\rho}\ra
 $ &\tiny 446 &\tiny 407 &\tiny 256 \\
 \tiny$ \la u^{\mu}u^{\nu}u_{\mu}u^{\lambda}\ra\la h_{\nu\lambda}\ra
 $ &\tiny 291 &\tiny 263 &\tiny 167
 & \tiny$ \la u^{\mu}\ra\la u^{\nu}\chip h_{\mu\nu}\ra
 +h.c.$ &\tiny 343 &\tiny 311 &\tiny 195
 & \tiny$ \la f_{+}^{\mu\nu}u_{\mu}u_{\nu}\ra\la\chim\ra
 $ &\tiny 395 &\tiny 358 &
 & \tiny$i \epsilon^{\mu\nu\lambda\rho}\la f_{+\mu\nu}u^{\sigma}u_{\lambda}u_{\rho}u_{\sigma}\ra
 $ &\tiny 447 &\tiny 408 &\tiny 257 \\
\hline\hline
 \end{longtable}}

 {\tabcolsep 1pt\renewcommand{\arraystretch}{0.4}
 \begin{longtable}{l|c|c|c|c|c|c|c|c|c|c|c|c|l|c|c|c|c|c|c|c|c|c|c|c|c}
 \caption{\label{p6tab3}The $p^6$ order results in U groups and SU groups, with $t=0,\thetah\neq0$ and the even parity.
 The numbers are the sequence numbers in each case.
 The details can be found in Section \ref{p6result}.}\\

 \hline\hline & \multicolumn{3}{c|}{\zh U(3)}
 & \multicolumn{3}{c|}{\zh SU(3)$_\mathrm{I}$}
 & \multicolumn{3}{c|}{\zh SU(3)$_\mathrm{II}$}
 & \multicolumn{3}{c|}{\zh SU(3)$_\mathrm{III}$}
 & & \multicolumn{3}{c|}{\zh U(3)}
 & \multicolumn{3}{c|}{\zh SU(3)$_\mathrm{I}$}
 & \multicolumn{3}{c|}{\zh SU(3)$_\mathrm{II}$}
 & \multicolumn{3}{c}{\zh SU(3)$_\mathrm{III}$} \\
 \hline \zh operators & \zh $n$ & \zh $3$ & \zh $2$ & \zh $n$ & \zh $3$ & \zh $2$ & \zh $n$ & \zh $3$ & \zh $2$ & \zh $n$ & \zh $3$ & \zh $2$
 & \zh operators & \zh $n$ & \zh $3$ & \zh $2$ & \zh $n$ & \zh $3$ & \zh $2$ & \zh $n$ & \zh $3$ & \zh $2$ & \zh $n$ & \zh $3$ & \zh $2$\\
 \hline\endfirsthead

 \hline\hline \zh operators & \zh $n$ & \zh $3$ & \zh $2$ & \zh $n$ & \zh $3$ & \zh $2$ & \zh $n$ & \zh $3$ & \zh $2$ & \zh $n$ & \zh $3$ & \zh $2$
 & \zh operators & \zh $n$ & \zh $3$ & \zh $2$ & \zh $n$ & \zh $3$ & \zh $2$ & \zh $n$ & \zh $3$ & \zh $2$ & \zh $n$ & \zh $3$ & \zh $2$\\
 \hline\endhead

 \hline\hline 
 \endfoot

 \hline\endlastfoot

 \tiny$i \la u^{\mu}u_{\mu}u^{\nu}u_{\nu}u^{\lambda}\ra\nabla_{\lambda}\thetah
 $ &\tiny 1 &\tiny 1 &\tiny 1 &\tiny 1 &\tiny 1 &  &\tiny 1 &\tiny 1 &\tiny 1 &\tiny 1 &\tiny 1 &
 & \tiny$ \la u^{\mu}\ra\la u_{\mu}\ra\la u^{\nu}u^{\lambda}\ra\nabla_{\nu}\thetah \nabla_{\lambda}\thetah
 $ &\tiny 133 &\tiny 128 &  &  &  &  &\tiny 127 &\tiny 122 &  &  &  &  \\
 \tiny$i \la u^{\mu}u_{\mu}u^{\nu}u^{\lambda}u_{\nu}\ra\nabla_{\lambda}\thetah
 $ &\tiny 2 &\tiny 2 &\tiny 2 &\tiny 2 &\tiny 2 &  &\tiny 2 &\tiny 2 &\tiny 2 &\tiny 2 &\tiny 2 &
 & \tiny$ \la u^{\mu}\ra\la u^{\nu}\ra\la u_{\mu}u^{\lambda}\ra\nabla_{\nu}\thetah \nabla_{\lambda}\thetah
 $ &\tiny 134 &\tiny 129 &  &  &  &  &\tiny 128 &\tiny 123 &  &  &  &  \\
 \tiny$i \la u^{\mu}u^{\nu}u_{\mu}u_{\nu}u^{\lambda}\ra\nabla_{\lambda}\thetah
 $ &\tiny 3 &\tiny 3 &\tiny 3 &\tiny 3 &\tiny 3 &  &\tiny 3 &\tiny 3 &\tiny 3 &\tiny 3 &\tiny 3 &
 & \tiny$ \la u^{\mu}\ra\la u^{\nu}\ra\la u^{\lambda}u_{\lambda}\ra\nabla_{\mu}\thetah \nabla_{\nu}\thetah
 $ &\tiny 135 &\tiny 130 &  &  &  &  &\tiny 129 &\tiny 124 &  &  &  &  \\
 \tiny$i \la u^{\mu}\ra\la u_{\mu}u^{\nu}u_{\nu}u^{\lambda}\ra\nabla_{\lambda}\thetah
 +h.c.$ &\tiny 4 &\tiny 4 &\tiny 4 &  &  &  &\tiny 4 &\tiny 4 &\tiny 4 &  &  &
 & \tiny$ \la u^{\mu}\ra\la u_{\mu}\ra\la u^{\nu}\ra\la u_{\nu}\ra\nabla^{\lambda}\thetah \nabla_{\lambda}\thetah
 $ &\tiny 136 &  &  &  &  &  &\tiny 130 &  &  &  &  &  \\
 \tiny$i \la u^{\mu}u_{\mu}\ra\la u^{\nu}u_{\nu}u^{\lambda}\ra\nabla_{\lambda}\thetah
 $ &\tiny 5 &\tiny 5 &\tiny 5 &\tiny 4 &\tiny 4 &  &\tiny 5 &\tiny 5 &\tiny 5 &\tiny 4 &\tiny 4 &
 & \tiny$ \la u^{\mu}\ra\la u_{\mu}\ra\la u^{\nu}\ra\la u^{\lambda}\ra\nabla_{\nu}\thetah \nabla_{\lambda}\thetah
 $ &\tiny 137 &  &  &  &  &  &\tiny 131 &  &  &  &  &  \\
\hline\hline
 \end{longtable}}

 {\tabcolsep 1pt\renewcommand{\arraystretch}{0.4}
 \renewcommand{\oddsidemargin}{-1cm}
 \begin{longtable}{@{\hspace{-1cm}}l|c|c|c|l|c|c|c|l|c|c|c|l|c|c|c}
 \caption{\label{p6tab4}The $p^6$ order results in U groups and SU groups, with $t=0$, $\thetah\neq0$ and the odd parity.
 The numbers are the sequence numbers in each case.
 The details can be found in Section \ref{p6result}.}\\

 \hline\hline & \multicolumn{3}{c|}{\zh U(3)}
 & & \multicolumn{3}{c|}{\zh U(3)}
 & & \multicolumn{3}{c|}{\zh U(3)}
 & & \multicolumn{3}{c}{\zh U(3)} \\
 \hline \zh operators & \zh $n$ & \zh $3$ & \zh $2$ &\zh operators & \zh $n$ & \zh $3$ & \zh $2$
 &\zh operators & \zh $n$ & \zh $3$ & \zh $2$ &\zh operators & \zh $n$ & \zh $3$ & \zh $2$ \\
 \hline\endfirsthead

 \hline\hline \zh operators & \zh $n$ & \zh $3$ & \zh $2$ &\zh operators & \zh $n$ & \zh $3$ & \zh $2$
 &\zh operators & \zh $n$ & \zh $3$ & \zh $2$ &\zh operators & \zh $n$ & \zh $3$ & \zh $2$ \\
 \hline\endhead

 \hline\hline 
 \endfoot

 \hline\endlastfoot

 \tiny$i \la u^{\mu}u_{\mu}u^{\nu}{f_{-\nu}}^{\lambda}\ra\nabla_{\lambda}\thetah
 +h.c.$ &\tiny 262 &\tiny 254 &\tiny 205
 & \tiny$ \la\nabla^{\mu}{f_{+\mu}}^{\nu}u_{\nu}u^{\lambda}\ra\nabla_{\lambda}\thetah
 +h.c.$ &\tiny 314 &\tiny 303 &\tiny 241
 & \tiny$ \la u^{\mu}\ra\la u^{\nu}{f_{-\mu}}^{\lambda}\ra\nabla_{\nu}\thetah \nabla_{\lambda}\thetah
 $ &\tiny 366 &\tiny 354 &\tiny 283
 & \tiny$i \la F_{L}^{\mu\nu}D^{\lambda}F_{L\mu\nu}\ra\nabla_{\lambda}\thetah
 +h.c.$ &\tiny 418 &\tiny 406 &\tiny 329 \\
 \tiny$i \la u^{\mu}u^{\nu}u_{\mu}{f_{-\nu}}^{\lambda}\ra\nabla_{\lambda}\thetah
 $ &\tiny 263 &\tiny 255 &\tiny 206
 & \tiny$i \la u^{\mu}\chip{f_{-\mu}}^{\nu}\ra\nabla_{\nu}\thetah
 +h.c.$ &\tiny 315 &\tiny 304 &\tiny 242
 & \tiny$ \la u^{\mu}u^{\nu}\ra\la{f_{-\mu}}^{\lambda}\ra\nabla_{\nu}\thetah \nabla_{\lambda}\thetah
 $ &\tiny 367 &\tiny 355 &\tiny 284
 & \tiny$i \la D^{\mu}\chi \chi^{\dag}\ra\nabla_{\mu}\thetah
 +h.c.$ &\tiny 419 &\tiny 407 &\tiny 330 \\
 \tiny$i \la u^{\mu}u^{\nu}u^{\lambda}f_{-\mu\nu}\ra\nabla_{\lambda}\thetah
 +h.c.$ &\tiny 264 &\tiny 256 &\tiny 207
 & \tiny$i \la u^{\mu}\chip{h_{\mu}}^{\nu}\ra\nabla_{\nu}\thetah
 +h.c.$ &\tiny 316 &\tiny 305 &\tiny 243
 & \tiny$ \la u^{\mu}\ra\la u^{\nu}{f_{-\nu}}^{\lambda}\ra\nabla_{\mu}\thetah \nabla_{\lambda}\thetah
 $ &\tiny 368 &\tiny 356 &\tiny 285
 & \tiny$i \nabla^{\mu}\nabla^{\nu}\thetah \nabla_{\nu}\nabla^{\lambda}\thetah \nabla_{\lambda}\nabla_{\mu}\thetah
 $ &\tiny 420 &\tiny 408 &\tiny 331 \\
 \tiny$i \la u^{\mu}u_{\mu}u^{\nu}{h_{\nu}}^{\lambda}\ra\nabla_{\lambda}\thetah
 +h.c.$ &\tiny 265 &\tiny 257 &\tiny 208
 & \tiny$i \la u^{\mu}u_{\mu}\chi_{+}^{\nu}\ra\nabla_{\nu}\thetah
 $ &\tiny 317 &\tiny 306 &\tiny 244
 & \tiny$ \la u^{\mu}\ra\la u_{\mu}h^{\nu\lambda}\ra\nabla_{\nu}\thetah \nabla_{\lambda}\thetah
 $ &\tiny 369 &\tiny 357 &\tiny 286
 & \tiny$i \nabla^{\mu}\nabla_{\mu}\thetah \nabla^{\nu}\nabla^{\lambda}\thetah \nabla_{\lambda}\nabla_{\nu}\thetah
 $ &\tiny 421 &\tiny 409 &\tiny 332 \\
\hline\hline
 \end{longtable}}

 {\tabcolsep 1pt\renewcommand{\arraystretch}{0.4}
 \begin{longtable}{l|c|c|c|c|c|c|c|c|c|c|c|c|l|c|c|c|c|c|c|c|c|c|c|c|c}
 \caption{\label{p6tab5}The $p^6$ order results in U groups and SU groups, with $t\neq0$, $\thetah=0$ and the even parity.
 The numbers are the sequence numbers in each case.
 The details can be found in Section \ref{p6result}.}\\

 \hline\hline & \multicolumn{3}{c|}{\zh U(3)}
 & \multicolumn{3}{c|}{\zh SU(3)$_\mathrm{I}$}
 & \multicolumn{3}{c|}{\zh SU(3)$_\mathrm{II}$}
 & \multicolumn{3}{c|}{\zh SU(3)$_\mathrm{III}$}
 & & \multicolumn{3}{c|}{\zh U(3)}
 & \multicolumn{3}{c|}{\zh SU(3)$_\mathrm{I}$}
 & \multicolumn{3}{c|}{\zh SU(3)$_\mathrm{II}$}
 & \multicolumn{3}{c}{\zh SU(3)$_\mathrm{III}$} \\
 \hline \zh operators & \zh $n$ & \zh $3$ & \zh $2$ & \zh $n$ & \zh $3$ & \zh $2$ & \zh $n$ & \zh $3$ & \zh $2$ & \zh $n$ & \zh $3$ & \zh $2$
 & \zh operators & \zh $n$ & \zh $3$ & \zh $2$ & \zh $n$ & \zh $3$ & \zh $2$ & \zh $n$ & \zh $3$ & \zh $2$ & \zh $n$ & \zh $3$ & \zh $2$\\
 \hline\endfirsthead

 \hline\hline \zh operators & \zh $n$ & \zh $3$ & \zh $2$ & \zh $n$ & \zh $3$ & \zh $2$ & \zh $n$ & \zh $3$ & \zh $2$ & \zh $n$ & \zh $3$ & \zh $2$
 & \zh operators & \zh $n$ & \zh $3$ & \zh $2$ & \zh $n$ & \zh $3$ & \zh $2$ & \zh $n$ & \zh $3$ & \zh $2$ & \zh $n$ & \zh $3$ & \zh $2$\\
 \hline\endhead

 \hline\hline 
 \endfoot

 \hline\endlastfoot

 \tiny$i \la u^{\mu}u_{\mu}u^{\nu}u^{\lambda}t_{+\nu\lambda}\ra
 +h.c.$ &\tiny 1 &\tiny 1 &\tiny 1 &\tiny 1 &\tiny 1 &\tiny 1 &\tiny 1 &\tiny 1 &\tiny 1 &\tiny 1 &\tiny 1 &\tiny 1
 & \tiny$ \la u^{\mu}\ra\la u^{\nu}\ra\la{t_{+\mu}}^{\lambda}t_{+\nu\lambda}\ra
 $ &\tiny 81 &\tiny 77 &  &  &  &  &\tiny 69 &\tiny 65 &  &  &  &  \\
 \tiny$i \la u^{\mu}u^{\nu}u_{\mu}u^{\lambda}t_{+\nu\lambda}\ra
 +h.c.$ &\tiny 2 &\tiny 2 &\tiny 2 &\tiny 2 &\tiny 2 &\tiny 2 &\tiny 2 &\tiny 2 &\tiny 2 &\tiny 2 &\tiny 2 &\tiny 2
 & \tiny$ \la u^{\mu}\ra\la u^{\nu}{t_{+\mu}}^{\lambda}\ra\la t_{+\nu\lambda}\ra
 $ &\tiny 82 &\tiny 78 &  &  &  &  &\tiny 70 &\tiny 66 &  &  &  &  \\
 \tiny$i \la u^{\mu}u^{\nu}u_{\nu}u^{\lambda}t_{+\mu\lambda}\ra
 $ &\tiny 3 &\tiny 3 &\tiny 3 &\tiny 3 &\tiny 3 &  &\tiny 3 &\tiny 3 &\tiny 3 &\tiny 3 &\tiny 3 &
 & \tiny$ \la u^{\mu}u^{\nu}\ra\la{t_{+\mu}}^{\lambda}\ra\la t_{+\nu\lambda}\ra
 $ &\tiny 83 &\tiny 79 &  &\tiny 57 &  &  &\tiny 71 &\tiny 67 &  &\tiny 50 &  &  \\
 \tiny$i \la u^{\mu}u^{\nu}u^{\lambda}u_{\mu}t_{+\nu\lambda}\ra
 $ &\tiny 4 &\tiny 4 &\tiny 4 &\tiny 4 &\tiny 4 &  &\tiny 4 &\tiny 4 &\tiny 4 &\tiny 4 &\tiny 4 &
 & \tiny$ \la u^{\mu}\ra\la u^{\nu}{t_{+\nu}}^{\lambda}\ra\la t_{+\mu\lambda}\ra
 $ &\tiny 84 &\tiny 80 &  &  &  &  &\tiny 72 &\tiny 68 &  &  &  &  \\
 \tiny$i \la u^{\mu}\ra\la u_{\mu}u^{\nu}u^{\lambda}t_{+\nu\lambda}\ra
 +h.c.$ &\tiny 5 &\tiny 5 &\tiny 5 &  &  &  &\tiny 5 &\tiny 5 &\tiny 5 &  &  &
 & \tiny$ \la u^{\mu}\ra\la u_{\mu}\ra\la t_{+}^{\nu\lambda}\ra\la t_{+\nu\lambda}\ra
 $ &\tiny 85 &  &  &  &  &  &\tiny 73 &  &  &  &  &  \\
 \tiny$i \la u^{\mu}u_{\mu}\ra\la u^{\nu}u^{\lambda}t_{+\nu\lambda}\ra
 $ &\tiny 6 &\tiny 6 &  &\tiny 5 &\tiny 5 &  &\tiny 6 &\tiny 6 &  &\tiny 5 &\tiny 5 &
 & \tiny$ \la u^{\mu}\ra\la u^{\nu}\ra\la{t_{+\mu}}^{\lambda}\ra\la t_{+\nu\lambda}\ra
 $ &\tiny 86 &  &  &  &  &  &\tiny 74 &  &  &  &  &  \\
\hline\hline
 \end{longtable}}

 {\tabcolsep 1pt\renewcommand{\arraystretch}{0.4}
 \renewcommand{\oddsidemargin}{-1cm}
 \begin{longtable}{@{\hspace{-1cm}}l|c|c|c|l|c|c|c|l|c|c|c|l|c|c|c}
 \caption{\label{p6tab6}The $p^6$ order results in U groups and SU groups, with $t\neq0$, $\thetah=0$ and the odd parity.
 The numbers are the sequence numbers in each case.
 The details can be found in Section \ref{p6result}.}\\

 \hline\hline & \multicolumn{3}{c|}{\zh U(3)}
 & & \multicolumn{3}{c|}{\zh U(3)}
 & & \multicolumn{3}{c|}{\zh U(3)}
 & & \multicolumn{3}{c}{\zh U(3)} \\
 \hline \zh operators & \zh $n$ & \zh $3$ & \zh $2$ &\zh operators & \zh $n$ & \zh $3$ & \zh $2$
 &\zh operators & \zh $n$ & \zh $3$ & \zh $2$ &\zh operators & \zh $n$ & \zh $3$ & \zh $2$ \\
 \hline\endfirsthead

 \hline\hline \zh operators & \zh $n$ & \zh $3$ & \zh $2$ &\zh operators & \zh $n$ & \zh $3$ & \zh $2$
 &\zh operators & \zh $n$ & \zh $3$ & \zh $2$ &\zh operators & \zh $n$ & \zh $3$ & \zh $2$ \\
 \hline\endhead

 \hline\hline 
 \endfoot

 \hline\endlastfoot

 \tiny$i \la u^{\mu}u_{\mu}f_{-}^{\nu\lambda}t_{+\nu\lambda}\ra
 +h.c.$ &\tiny 160 &\tiny 154 &\tiny 109
 & \tiny$ \la u^{\mu}u^{\nu}\chim t_{+\mu\nu}\ra
 +h.c.$ &\tiny 198 &\tiny 192 &\tiny 136
 & \tiny$ \la u^{\mu}\ra\la u^{\nu}\ra\la u_{\mu}u^{\lambda}t_{-\nu\lambda}\ra
 +h.c.$ &\tiny 236 &  &
 & \tiny$ \la u^{\mu}u^{\nu}\chip t_{-\mu\nu}\ra
 +h.c.$ &\tiny 274 &\tiny 264 &\tiny 185 \\
 \tiny$i \la u^{\mu}u^{\nu}{f_{-\mu}}^{\lambda}t_{+\nu\lambda}\ra
 +h.c.$ &\tiny 161 &\tiny 155 &\tiny 110
 & \tiny$ \la u^{\mu}\chim u^{\nu}t_{+\mu\nu}\ra
 $ &\tiny 199 &\tiny 193 &\tiny 137
 & \tiny$ \la f_{-}^{\mu\nu}{f_{-\mu}}^{\lambda}t_{-\nu\lambda}\ra
 $ &\tiny 237 &\tiny 229 &\tiny 160
 & \tiny$ \la u^{\mu}\chip u^{\nu}t_{-\mu\nu}\ra
 $ &\tiny 275 &\tiny 265 &\tiny 186 \\
 \tiny$i \la u^{\mu}u^{\nu}{f_{-\nu}}^{\lambda}t_{+\mu\lambda}\ra
 +h.c.$ &\tiny 162 &\tiny 156 &\tiny 111
 & \tiny$ \la u^{\mu}\ra\la u^{\nu}\chim t_{+\mu\nu}\ra
 +h.c.$ &\tiny 200 &\tiny 194 &\tiny 138
 & \tiny$ \la h^{\mu\nu}{f_{-\mu}}^{\lambda}t_{-\nu\lambda}\ra
 +h.c.$ &\tiny 238 &\tiny 230 &\tiny 161
 & \tiny$ \la u^{\mu}\ra\la u^{\nu}\chip t_{-\mu\nu}\ra
 +h.c.$ &\tiny 276 &\tiny 266 &\tiny 187 \\
 \tiny$i \la u^{\mu}{f_{-\mu}}^{\nu}u^{\lambda}t_{+\nu\lambda}\ra
 +h.c.$ &\tiny 163 &\tiny 157 &\tiny 112
 & \tiny$ \la u^{\mu}u^{\nu}\chim\ra\la t_{+\mu\nu}\ra
 $ &\tiny 201 &\tiny 195 &
 & \tiny$ \la h^{\mu\nu}{h_{\mu}}^{\lambda}t_{-\nu\lambda}\ra
 $ &\tiny 239 &\tiny 231 &\tiny 162
 & \tiny$ \la u^{\mu}u^{\nu}\chip\ra\la t_{-\mu\nu}\ra
 $ &\tiny 277 &\tiny 267 &  \\
 \tiny$i \la u^{\mu}u^{\nu}{h_{\mu}}^{\lambda}t_{+\nu\lambda}\ra
 +h.c.$ &\tiny 164 &\tiny 158 &\tiny 113
 & \tiny$ \la u^{\mu}u^{\nu}t_{+\mu\nu}\ra\la\chim\ra
 $ &\tiny 202 &\tiny 196 &
 & \tiny$ \la u^{\mu}\nabla_{\mu}f_{-}^{\nu\lambda}t_{-\nu\lambda}\ra
 +h.c.$ &\tiny 240 &\tiny 232 &\tiny 163
 & \tiny$ \la u^{\mu}u^{\nu}t_{-\mu\nu}\ra\la\chip\ra
 $ &\tiny 278 &\tiny 268 &  \\
\hline\hline
 \end{longtable}}

 {\tabcolsep 1pt\renewcommand{\arraystretch}{0.4}
 \begin{longtable}{l|c|c|c|c|c|c|c|c|c|c|c|c|l|c|c|c|c|c|c|c|c|c|c|c|c}
 \caption{\label{p6tab7}The $p^6$ order results in U groups and SU groups, with $t\neq0$, $\thetah\neq0$ and the even parity.
 The numbers are the sequence numbers in each case.
 The details can be found in Section \ref{p6result}.}\\

 \hline\hline & \multicolumn{3}{c|}{\zh U(3)}
 & \multicolumn{3}{c|}{\zh SU(3)$_\mathrm{I}$}
 & \multicolumn{3}{c|}{\zh SU(3)$_\mathrm{II}$}
 & \multicolumn{3}{c|}{\zh SU(3)$_\mathrm{III}$}
 & & \multicolumn{3}{c|}{\zh U(3)}
 & \multicolumn{3}{c|}{\zh SU(3)$_\mathrm{I}$}
 & \multicolumn{3}{c|}{\zh SU(3)$_\mathrm{II}$}
 & \multicolumn{3}{c}{\zh SU(3)$_\mathrm{III}$} \\
 \hline \zh $m$ & \zh $n$ & \zh $3$ & \zh $2$ & \zh $n$ & \zh $3$ & \zh $2$ & \zh $n$ & \zh $3$ & \zh $2$ & \zh $n$ & \zh $3$ & \zh $2$
 & \zh $m$ & \zh $n$ & \zh $3$ & \zh $2$ & \zh $n$ & \zh $3$ & \zh $2$ & \zh $n$ & \zh $3$ & \zh $2$ & \zh $n$ & \zh $3$ & \zh $2$\\
 \hline\endfirsthead

 \hline\hline \zh $m$ & \zh $n$ & \zh $3$ & \zh $2$ & \zh $n$ & \zh $3$ & \zh $2$ & \zh $n$ & \zh $3$ & \zh $2$ & \zh $n$ & \zh $3$ & \zh $2$
 & \zh $m$ & \zh $n$ & \zh $3$ & \zh $2$ & \zh $n$ & \zh $3$ & \zh $2$ & \zh $n$ & \zh $3$ & \zh $2$ & \zh $n$ & \zh $3$ & \zh $2$\\
 \hline\endhead

 \hline\hline 
 \endfoot

 \hline\endlastfoot

 \tiny$ \la u^{\mu}u_{\mu}u^{\nu}{t_{+\nu}}^{\lambda}\ra\nabla_{\lambda}\thetah
 +h.c.$ &\tiny 1 &\tiny 1 &\tiny 1 &\tiny 1 &\tiny 1 &  &\tiny 1 &\tiny 1 &\tiny 1 &\tiny 1 &\tiny 1 &
 & \tiny$i \la u^{\mu}{t_{+\mu}}^{\nu}\ra\la{t_{+\nu}}^{\lambda}\ra\nabla_{\lambda}\thetah
 $ &\tiny 33 &\tiny 33 &\tiny 27 &\tiny 23 &\tiny 23 &\tiny 17 &\tiny 25 &\tiny 25 &\tiny 19 &\tiny 18 &\tiny 18 &\tiny 12 \\
 \tiny$ \la u^{\mu}u^{\nu}u^{\lambda}t_{+\mu\nu}\ra\nabla_{\lambda}\thetah
 +h.c.$ &\tiny 2 &\tiny 2 &\tiny 2 &\tiny 2 &\tiny 2 &\tiny 1 &\tiny 2 &\tiny 2 &\tiny 2 &\tiny 2 &\tiny 2 &\tiny 1
 & \tiny$i \la u^{\mu}t_{+}^{\nu\lambda}\ra\la t_{+\mu\nu}\ra\nabla_{\lambda}\thetah
 $ &\tiny 34 &\tiny 34 &\tiny 28 &\tiny 24 &\tiny 24 &  &\tiny 26 &\tiny 26 &\tiny 20 &\tiny 19 &\tiny 19 &  \\
 \tiny$ \la u^{\mu}u^{\nu}u^{\lambda}t_{+\mu\lambda}\ra\nabla_{\nu}\thetah
 $ &\tiny 3 &\tiny 3 &\tiny 3 &\tiny 3 &\tiny 3 &  &\tiny 3 &\tiny 3 &\tiny 3 &\tiny 3 &\tiny 3 &
 & \tiny$i \la u^{\mu}\ra\la t_{+}^{\nu\lambda}t_{+\nu\lambda}\ra\nabla_{\mu}\thetah
 $ &\tiny 35 &\tiny 35 &\tiny 29 &  &  &  &\tiny 27 &\tiny 27 &\tiny 21 &  &  &  \\
 \tiny$ \la u^{\mu}\ra\la u_{\mu}u^{\nu}{t_{+\nu}}^{\lambda}\ra\nabla_{\lambda}\thetah
 +h.c.$ &\tiny 4 &\tiny 4 &  &  &  &  &\tiny 4 &\tiny 4 &  &  &  &
 & \tiny$i \la u^{\mu}t_{+}^{\nu\lambda}\ra\la t_{+\nu\lambda}\ra\nabla_{\mu}\thetah
 $ &\tiny 36 &\tiny 36 &\tiny 30 &\tiny 25 &\tiny 25 &  &\tiny 28 &\tiny 28 &\tiny 22 &\tiny 20 &\tiny 20 &  \\
 \tiny$ \la u^{\mu}\ra\la u^{\nu}u^{\lambda}t_{+\mu\nu}\ra\nabla_{\lambda}\thetah
 +h.c.$ &\tiny 5 &\tiny 5 &\tiny 4 &  &  &  &\tiny 5 &\tiny 5 &\tiny 4 &  &  &
 & \tiny$i \la u^{\mu}\ra\la{t_{+\mu}}^{\nu}\ra\la{t_{+\nu}}^{\lambda}\ra\nabla_{\lambda}\thetah
 $ &\tiny 37 &\tiny 37 &  &  &  &  &\tiny 29 &\tiny 29 &  &  &  &  \\
\hline\hline
 \end{longtable}}

  {\tabcolsep 1pt\renewcommand{\arraystretch}{0.4}
 \renewcommand{\oddsidemargin}{-1cm}
 \begin{longtable}{@{\hspace{-1cm}}l|c|c|c|l|c|c|c|l|c|c|c|l|c|c|c}
 \caption{\label{p6tab8}The $p^6$ order results in U groups and SU groups, with $t\neq0$, $\thetah\neq0$ and the odd parity.
 The numbers are the sequence numbers in each case.
 The details can be found in Section \ref{p6result}.}\\

 \hline\hline & \multicolumn{3}{c|}{\zh U(3)}
 & & \multicolumn{3}{c|}{\zh U(3)}
 & & \multicolumn{3}{c|}{\zh U(3)}
 & & \multicolumn{3}{c}{\zh U(3)} \\
 \hline \zh operators & \zh $n$ & \zh $3$ & \zh $2$ &\zh operators & \zh $n$ & \zh $3$ & \zh $2$
 &\zh operators & \zh $n$ & \zh $3$ & \zh $2$ &\zh operators & \zh $n$ & \zh $3$ & \zh $2$ \\
 \hline\endfirsthead

 \hline\hline \zh operators & \zh $n$ & \zh $3$ & \zh $2$ &\zh operators & \zh $n$ & \zh $3$ & \zh $2$
 &\zh operators & \zh $n$ & \zh $3$ & \zh $2$ &\zh operators & \zh $n$ & \zh $3$ & \zh $2$ \\
 \hline\endhead

 \hline\hline 
 \endfoot

 \hline\endlastfoot

 \tiny$ \la u^{\mu}{f_{-\mu}}^{\nu}{t_{+\nu}}^{\lambda}\ra\nabla_{\lambda}\thetah
 +h.c.$ &\tiny 65 &\tiny 65 &\tiny 57
 & \tiny$i \la u^{\mu}\chim{t_{+\mu}}^{\nu}\ra\nabla_{\nu}\thetah
 +h.c.$ &\tiny 80 &\tiny 80 &\tiny 72
 & \tiny$ \la f_{+}^{\mu\nu}u^{\lambda}t_{-\mu\lambda}\ra\nabla_{\nu}\thetah
 +h.c.$ &\tiny 95 &\tiny 95 &\tiny 84
 & \tiny$ \la u^{\mu}u^{\nu}{t_{-\mu}}^{\lambda}\ra\nabla_{\nu}\thetah \nabla_{\lambda}\thetah
 +h.c.$ &\tiny 110 &\tiny 110 &\tiny 96 \\
 \tiny$ \la u^{\mu}f_{-}^{\nu\lambda}t_{+\mu\nu}\ra\nabla_{\lambda}\thetah
 +h.c.$ &\tiny 66 &\tiny 66 &\tiny 58
 & \tiny$i \la t_{+}^{\mu\nu}t_{+\mu\nu}\ra\nabla^{\lambda}\nabla_{\lambda}\thetah
 $ &\tiny 81 &\tiny 81 &\tiny 73
 & \tiny$ \la f_{+}^{\mu\nu}\ra\la u_{\mu}{t_{-\nu}}^{\lambda}\ra\nabla_{\lambda}\thetah
 $ &\tiny 96 &\tiny 96 &\tiny 85
 & \tiny$i \la f_{+}^{\mu\nu}t_{-\mu\nu}\ra\nabla^{\lambda}\thetah \nabla_{\lambda}\thetah
 $ &\tiny 111 &\tiny 111 &\tiny 97 \\
 \tiny$ \la u^{\mu}f_{-}^{\nu\lambda}t_{+\nu\lambda}\ra\nabla_{\mu}\thetah
 +h.c.$ &\tiny 67 &\tiny 67 &\tiny 59
 & \tiny$i \la\nabla^{\mu}{t_{+\mu}}^{\nu}{t_{+\nu}}^{\lambda}\ra\nabla_{\lambda}\thetah
 $ &\tiny 82 &\tiny 82 &\tiny 74
 & \tiny$ \la f_{+}^{\mu\nu}u_{\mu}\ra\la{t_{-\nu}}^{\lambda}\ra\nabla_{\lambda}\thetah
 $ &\tiny 97 &\tiny 97 &\tiny 86
 & \tiny$i \la f_{+}^{\mu\nu}{t_{-\mu}}^{\lambda}\ra\nabla_{\nu}\thetah \nabla_{\lambda}\thetah
 $ &\tiny 112 &\tiny 112 &\tiny 98 \\
 \tiny$ \la u^{\mu}{h_{\mu}}^{\nu}{t_{+\nu}}^{\lambda}\ra\nabla_{\lambda}\thetah
 +h.c.$ &\tiny 68 &\tiny 68 &\tiny 60
 & \tiny$i \la t_{+}^{\mu\nu}\ra\la t_{+\mu\nu}\ra\nabla^{\lambda}\nabla_{\lambda}\thetah
 $ &\tiny 83 &\tiny 83 &\tiny 75
 & \tiny$ \la f_{+}^{\mu\nu}{t_{-\mu}}^{\lambda}\ra\la u_{\nu}\ra\nabla_{\lambda}\thetah
 $ &\tiny 98 &\tiny 98 &\tiny 87
 & \tiny$i \la f_{+}^{\mu\nu}\ra\la t_{-\mu\nu}\ra\nabla^{\lambda}\thetah \nabla_{\lambda}\thetah
 $ &\tiny 113 &\tiny 113 &\tiny 99 \\
\hline\hline
 \end{longtable}}

 \bibliography{getcl4}
\end{document}